\documentclass[12pt, draftclsnofoot, onecolumn]{IEEEtran}

\usepackage[normalem]{ulem}

\usepackage{cite}
\usepackage[linesnumbered, algo2e, ruled,norelsize]{algorithm2e}
\usepackage{amsmath,amssymb,amsfonts}
\usepackage{algorithmic}
\usepackage{graphicx}
\usepackage{textcomp}
\usepackage{graphicx}
\usepackage{dsfont}
\usepackage{caption}
\usepackage{subcaption}
\usepackage{latexsym}
\usepackage{amsmath}
\usepackage[mathscr]{euscript}
\usepackage[linesnumbered, algo2e, ruled,norelsize]{algorithm2e}
\SetArgSty{textnormal}
\usepackage{threeparttable}
\usepackage{threeparttablex}
\usepackage{commath}
\usepackage{mathtools}
\usepackage{slashbox}
\usepackage{pict2e}
\usepackage{epstopdf}
\usepackage{verbatim}
\usepackage{subfloat}
\usepackage{tabu}
\usepackage{booktabs}
\usepackage{float}
\usepackage{amssymb}
\usepackage{url}
\usepackage{multirow}
\usepackage{graphicx}
\usepackage{mathptmx}
\usepackage{cite}
\usepackage{footnote}
\usepackage{array}
\usepackage[table]{xcolor}
\usepackage{tabularx}
\usepackage{ulem}
\usepackage{array}
\usepackage{slashbox}
\usepackage{lipsum}
\newcolumntype{P}[1]{>{\centering\arraybackslash}p{#1}}

\usepackage{amsthm} 
\makeatletter
\newcommand{\leftlabel}[1]{&&
  \refstepcounter{equation}\ltx@label{#1}%
  \tagform@{\theequation}&&}
\makeatletter

\def\BibTeX{{\rm B\kern-.05em{\sc i\kern-.025em b}\kern-.08em
    T\kern-.1667em\lower.7ex\hbox{E}\kern-.125emX}}

\begin{document}

\title{On the Design of Artificial Noise for Physical Layer Security in Visible Light Communication Channels with Clipping}

\author{Thanh~V.~Pham,
        Steve Hranilovic,
        and Susumu Ishihara 
\thanks{Thanh V. Pham and Susumu Ishihara are with the Department of Mathematical and Systems Engineering, Shizuoka University, Shizuoka, Japan (e-mail: pham.van.thanh@shizuoka.ac.jp, ishihara.susumu@shizuoka.ac.jp). 

Steve Hranilovic is with the Department of Electrical and Computer Engineering, McMaster University,
Hamilton L8S 4L8, Ontario, Canada (e-mail: hranilovic@mcmaster.ca).

Part of this paper has been accepted for presentation at the 2023 IEEE Consumer Communications and Networking Conference (2023 IEEE CCNC), Las Vegas, Nevada, USA, January 2023.}}
\maketitle
\begin{abstract}
Though visible light communication (VLC) systems are contained to a given room, improving their security is an important criterion in any practical deployment. In this paper, the design of artificial noise (AN)  to enhance physical layer security in VLC systems is studied in the context of input signals with no explicit amplitude constraint (e.g., multicarrier systems). In such systems, clipping is needed to constrain the input signals within the limited linear ranges of the LEDs.  However, this clipping process gives rise to non-linear clipping distortion, which must be incorporated into the AN design. To facilitate the solution of this problem, a sub-optimal design approach is presented using the Charnes-Cooper transformation and the convex-concave procedure (CCP). Then, a novel AN transmission scheme is proposed to reduce the impact of clipping distortion, thus improving the secrecy performance. The proposed scheme exploits the common structure of LED luminaries that they are often composed of several light-emitting chips. Capitalizing on this property, LED chips in each luminaire are divided into two groups driven by separate driver circuits. One group is  used to transmit the information-bearing signal, while the other group transmits the AN. Numerical results show that the clipping distortion significantly reduces the secrecy level, and using AN is advantageous over the no-AN scheme in improving the secrecy performance. Moreover, the proposed AN transmission scheme is shown to achieve considerable secrecy improvements compared with the traditional transmission approaches (e.g., about 1 bit/s/Hz improvement in the achievable secrecy rate when the standard deviation of the LEDs' modulating current is 0.25 A and the signal-to-interference-plus-noise ratio of the eavesdropper's received signal is limited to $0$ dB). 
\end{abstract}



\section{Introduction}
The continuously increasing demand for high-throughput wireless data transmissions necessitates the continued development of new communication technologies. In this context, visible light communications (VLC) is being considered a promising wireless technology to complement existing indoor wireless systems \cite{Memedi2021}. Aside from the ability to offer multi-Gbps connections \cite{Hu2022,Gutema2022}, research and development of VLC have also been motivated due to the increasing scarcity of radio frequency (RF) spectrum \cite{Kitsunezuka2012}, and the popularity of  light-emitting diodes (LEDs) for illumination \cite{doi/10.2760/759859}. 

As visible light is confined by opaque objects, VLC is expected to be more secure than RF. Nonetheless, there is still a security risk within the VLC system illumination areas due to the broadcast nature of the signal. The confidentiality of the transmitted information on such broadcast channels has been traditionally protected by means of key-based cryptographic algorithms employed at the upper layers of the OSI model. The security of these techniques is based on the assumption that the secret key can not be determined within a reasonable amount of time with the current computing power. However, the enormous progress in research and development of quantum computers, which are  expected to be exponentially faster than the best non-quantum one, may soon compromise the security of the classical cryptography \cite{Shor1994,Castelvecchi2022}. As a result, information-theoretic security approaches such as physical layer security (PLS) are of great interest in dealing with eavesdropping by malicious users having unbounded computing power. In this regard, recently, there has been an increasing interest in applying physical layer security (PLS) to enhance message confidentiality in VLC systems \cite{Blinowski2019,arfaoui2020}. Compared with RF systems, where an average power on the input is often considered, the input of VLC systems should be constrained by a peak power to ensure the proper operation of the LED transmitters. In the case of traditional wiretap channels consisting of a  transmitter, a legitimate user, and an eavesdropper, fundamental analyses on the secrecy capacity of systems  were presented in \cite{wang2018physical} and \cite{Wang2021} considering signal-independent and signal-dependent noise, respectively. Specifically, lower and upper bounds on the secrecy capacity were derived under an average optical intensity constraint and  both average and peak optical power constraints. Analyses on the secrecy outage probability (SOP) were also presented in \cite{zhao2018physical} for VLC systems employing non-orthogonal multiple access (NOMA). 

Since the primary function of VLC systems in indoor scenarios is illumination, multiple LED luminaires are typically deployed to meet the lighting requirement. On the one hand, this setup naturally gives rise to an optical attocell network where a user is served by a luminary at a given time \cite{Haas2013HighspeedWN}. Achievable secrecy rates considering different deployments of LED luminaires (i.e., hexagonal, square, Poisson point process, and hard-core point process) were compared in \cite{Chen2017}. On the other hand, multiple LED transmitters can cooperatively transmit signals to the user. In this case, the spatial degrees of freedom offered by spatially separated transmitters can be exploited in the form of precoding and artificial noise. Specifically, the use of precoding techniques to improve the PLS performance has been studied in \cite{mostafa2015physical,ma2016optimal,pham2017secrecy,arfaoui2018secrecy, Cho2018,Cho2020} and references therein. The benefit of AN in enhancing the achievable secrecy rate has also been intensively investigated in the literature \cite{Mostafa2014,Cho2019,Pham2021,Shen2016,Pham2018,pham2020energy}. It should be noted that AN is often utilized in combination with precoding to fully exploit the benefit of the spatial degrees of freedom available at the transmitters. Hence, such a combination scheme is considered in this paper. 

It is well-known that LEDs have specific input ranges over which the emitted optical power is linearly related to the amplitude of the input drive current. Due to this characteristic, previous works in \cite{Mostafa2014,Cho2019,Pham2021,Shen2016,Pham2018,pham2020energy} assumed that the AN and information-bearing signals are selected from amplitude-constrained constellations (e.g., pulse amplitude modulation - PAM) to ensure that the amplitude of the LED input drive current can be properly constrained within the LED linear range. However, the assumption of amplitude-constrained signals does not apply to multicarrier signals such as orthogonal frequency division multiplexing (OFDM), which are expected to be widely employed for high data-rates VLC systems \cite{Serafimovski2021}. Multicarrier signals can often have  large peak-to-average power ratios. Thus clipping is unavoidable in VLC OFDM systems as it is necessary to constrain the signals within the LED linear range. Nonetheless, clipping distortion is non-linear, which significantly complicates the design problems and renders solving the optimal solutions extremely difficult. 

The severity of clipping distortion is dependent on the amplitude of the LEDs' modulating signal and clipping levels. In the case of VLC, clipping levels are  fixed by the dynamic range of the LEDs. This leaves signal amplitude the sole factor influencing the clipping distortion. In previous studies, the LEDs' modulating signal is generated by combining the information-bearing and AN signals, which, in this paper, is referred to as the \textit{one-branch} AN transmission scheme.
Assuming that these two signals are independent and identically distributed, the amplitude of the combined modulating signal is then statistically larger than that of each individual one. The resulting clipping distortion is, therefore, more severe than it would be if LEDs were driven by only one signal. Nowadays, typical commercial LED luminaires are often composed of several LED chips to provide sufficient illumination. Capitalizing on this common structure, we propose a novel \textit{two-branch} AN transmission scheme by dividing LED chips in each luminary into two groups with separate circuits. The two groups are then driven separately by the information-bearing and AN signals. By doing so, since each LED chip is driven by either the information-bearing or the AN signal, the severity of clipping distortion can be reduced. It should be noted that the idea of grouping LED chips in a luminary has been explored in several studies for different purposes. For example, grouping LED chips to achieve digital-to-analog conversion (DAC) in the optimal domain was independently proposed in \cite{Armstrong2013}, and \cite{Fath2013}.  The authors in \cite{Mossaad2015} proposed a method to partition a wideband high peak-to-average power ratio (PAPR) OFDM signal into narrowband emissions, which are then used to drive groups of LED chips. The idea was also applied in \cite{Lin2019} to realize an optical domain non-orthogonal multiple access (OPD-NOMA) for multi-user VLC systems.  

In this paper, VLC systems with one legitimate user (i.e., Bob) and one eavesdropper (i.e., Eve) are examined where Eve's channel state information (CSI) can be either known or unknown at the transmitters. The design problems for the one-branch and two-branch systems are cast as a maximization of the signal-to-interference-plus-noise ratio (SINR) of Bob's received signal while limiting the SINR of Eve's eavesdropping signal below a predefined threshold. These problems, however, are shown to be non-convex, which renders solving optimal solutions difficult. Therefore, low-complexity sub-optimal designs using the Charnes-Cooper transformation and the convex-concave procedure (CCP) are presented. The contributions of this paper are specifically summarized as follows. 
\begin{itemize}
    \item Sub-optimal designs of the one-branch AN transmission scheme are presented for the case of known \cite{Pham2022} and unknown Eve's CSI at the transmitter. For the case of unknown Eve's CSI, the average Eve's SINR is first examined, which is, however, mathematically intractable. Thus, an alternative expression for the average Eve's SINR is taken into the AN design.   
    \item A novel two-branch AN transmission scheme to reduce the severity of clipping distortion is proposed. Sub-optimal designs of the proposed AN schemes are then studied for both scenarios of known and unknown Eve's CSI.  
    \item Extensive simulations are performed to thoroughly evaluate the impact of clipping distortion on the PLS performance of the one-branch AN transmission scheme and the improvement brought by the two-branch AN scheme. 
\end{itemize}

The balance of the paper is organized as follows. The system and channel models are described in Sec.~\ref{sec:sys-and-channel-model}. Sec.~\ref{sec:conventional-AN} presents  the designs of an AN scheme using a single branch OFDM transmitter with clipping. The proposed two-branch AN transmission scheme is explained in Sec.~\ref{sec:proposedAN}. Numerical results are discussed in Sec.~\ref{sec:numerical-discussions}. Finally, Sec.~\ref{sec:conclusion} concludes the paper with some remarks and future works. 

{\it{Notation}}: The following notations are used throughout the paper. Bold uppercase and lowercase letters (e.g., $\mathbf{H}$ and $\mathbf{h}$) represent matrices and vectors, respectively. $[\mathbf{H}]_{m, n}$ indicate the element at the $m$-th row and $n$-th column of $\mathbf{H}$ while $[\mathbf{h}]_n$ indicate the $n$-th element of $\mathbf{h}$. Also, $\mathbf{h}^T$ is the transpose of $\mathbf{h}$, $\mathbb{E}[\cdot]$ is the expected value operation, $\lVert\cdot\rVert$ is the Euclidean norm, and $\odot$ is the element-wise product operation. 

\section{System and Channel Models}
\label{sec:sys-and-channel-model}

\begin{figure}[ht]
    \centering
    \includegraphics[width = 10.0 cm, height = 7.0cm]{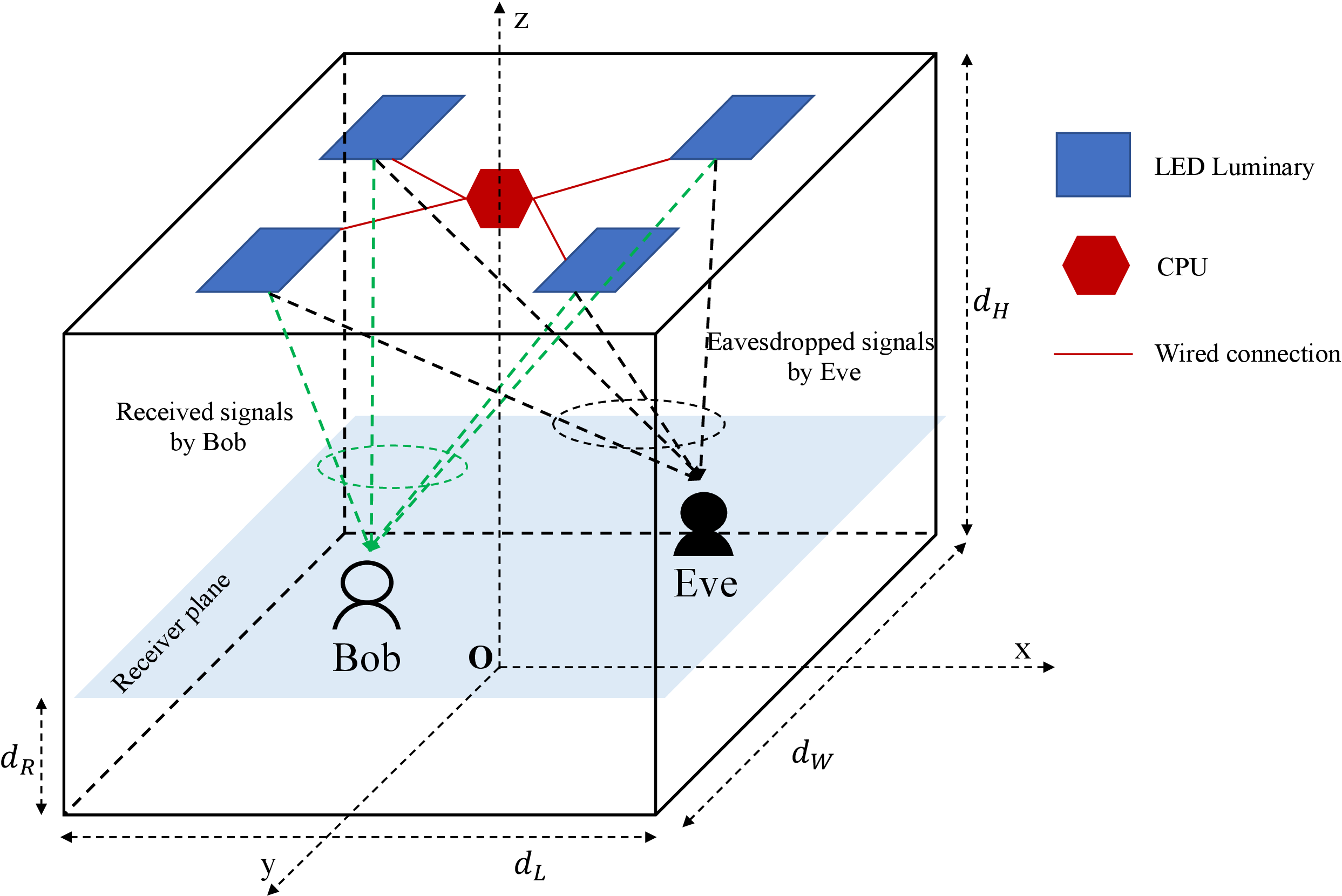}
    \caption{System Configuration.}
    \label{Fig0}
\end{figure}

We study in this paper a multi-transmitter VLC system that consists of a legitimate user (Bob) and an eavesdropper (Eve) who coexisted in a room of dimension $d_{\text{L}}~\text{(m)} \times d_{\text{W}}~\text{(m)} \times d_{\text{H}}~\text{(m)}$ as illustrated in Fig.~\ref{Fig0}.
There are $N_T$ LED luminaries connected to a Central Processing Unit (CPU) by wired connections. Assume that Bob and Eve are both equipped with a receiver having one photodiode (PD), and their receivers are on a receiver plane which is $d_{\text{R}}~\text{(m)}$ above the floor. For the sake of analytical simplicity, only the line-of-sight (LoS) propagation channel is considered. Let $\mathbf{h}_{\text{B}} = \begin{bmatrix}h_{1,\text{B}} & h_{2,\text{B}} & \hdots & h_{N_T, \text{B}}\end{bmatrix}^T$ be the channel vector of Bob where $h_{k, \text{B}}$ is the channel gain from the $k$-th LED luminary. Assume that the LED luminaries have Lambertian beam distribution and denote $m=\frac{-\log(2)}{\log(\cos\Phi_{1/2})}$ is the order of Lambertian emission where $\Phi_{1/2}$ is the semi-angle of half illuminance of the LEDs. Then, $h_{k, \text{B}}$ is given by
\begin{align}
    h_{k, \text{B}} = & \frac{(m+1)A_r}{2\pi l^2 _{k, \text{B}}}\cos^l(\phi_{k, \text{B}})T_s(\psi_{k, \text{B}})g(\psi_{k, \text{B}})\cos(\psi_{k, \text{B}}) \mathds{1}_{[0, \Psi]}(\psi_{k, \text{B}}),
    \label{Bob-channel-gain}
\end{align}
where $\mathds{1}_{[x, y]}(\cdot)$ denotes the indicator function, $A_r$ is the area of the PD and $l_{k, \text{B}}$ is the distance from the $k$-th luminary to Bob. $\psi_{k, \text{B}}$ is the angle of incidence and $\Psi$ is the optical field of view (FoV) of the PD. $T_s(\psi_{k, \text{B}})$ is the gain of the optical filter and $g_(\psi_{k, \text{B}})$ is the gain of the optical concentrator, which is given by
\begin{align}
    g(\psi_{k, \text{B}}) = \frac{\kappa^2}{\sin^2\Psi}\mathds{1}_{[0, \Psi]}(\psi_{k, \text{B}}),
    \label{concentrator-gain}
\end{align}
where $\kappa$ is the refractive index of the concentrator. 

Similarly, denote $\mathbf{h}_{\text{E}} = \begin{bmatrix}h_{1, \text{E}} & h_{2, \text{E}} & \cdots & h_{N_T, \text{E}}\end{bmatrix}^T$ as the channel vector of Eve, which can be calculated using \eqref{Bob-channel-gain} and \eqref{concentrator-gain}. In the case that Eve is an active eavesdropper, $\mathbf{h}_{\text{E}}$ can be known at the transmitting side due to feedback from Eve. In practical scenarios, Eve is often a passive malicious user who does not expose its presence to the transmitters. In this case, we assume that the instantaneous $\mathbf{h}_{\text{E}}$ is unknown to the transmitters. Instead, the average Eve's channel gain is calculated and used for the design and  analyses.  
\section{Designs of One-Branch AN Transmission with Clipping}
\label{sec:conventional-AN}
The IEEE 802.11bb standard advocates for the use of DC-biased optical (DCO) OFDM to support high data rates (up to 10 Gbps) transmission in VLC \cite{Serafimovski2021}. Unlike PAM and OOK signaling considered in previous studies, OFDM signals are inherently not subject to amplitude constraint and can have multiple high peaks. Hence, clipping is necessary to fit the signals into the LED linear range. 

Let $d$ and $z$ be the information-bearing and AN signals, respectively. In previous studies, $d$ and $z$ were either assumed to follow the uniform distribution \cite{Pham2018,pham2020energy} or the truncated Gaussian distribution \cite{Zaid2015,Cho2019, Arfaoui2019} over the normalized amplitude range $[-1,~1]$. In case of OFDM, when a large number of sub-carriers is used (e.g., 64 or more), the signals can be well approximated to be Gaussian and having domain over the real numbers. Hence, for the sake of analysis, let $d$ be a time-domain Gaussian information-bearing signal. Although the AN can be arbitrarily generated, to make the analysis tractable, let assume that $z$ is Gaussian distributed. Furthermore, without loss of generality, assume that $d$ and $z$ are independent zero-mean and have unit variance. At the $n$-th luminaire, denote $v_n$ and $w_n$ as the precoders of $d$ and $z$, respectively. The combined information-bearing and AN current is given by
\begin{align}
    x_n = v_nd + w_nz.
\end{align}
Since $d$ and $z$ are not amplitude-constrained, clippings are needed to limit the output currents within the LED's linear range, which is assumed to be over $[I_{\text{min}},~ I_{\text{max}}]$. Since $d\sim \mathcal{N}(0, 1)$ and $z \sim \mathcal{N}(0, 1)$, it follows that $x_n \sim \mathcal{N}\left(0, v^2_n + w^2_n\right)$.
\begin{figure}[ht]
    \centering
    \includegraphics[width = 0.65\textwidth, height = 7.0cm]{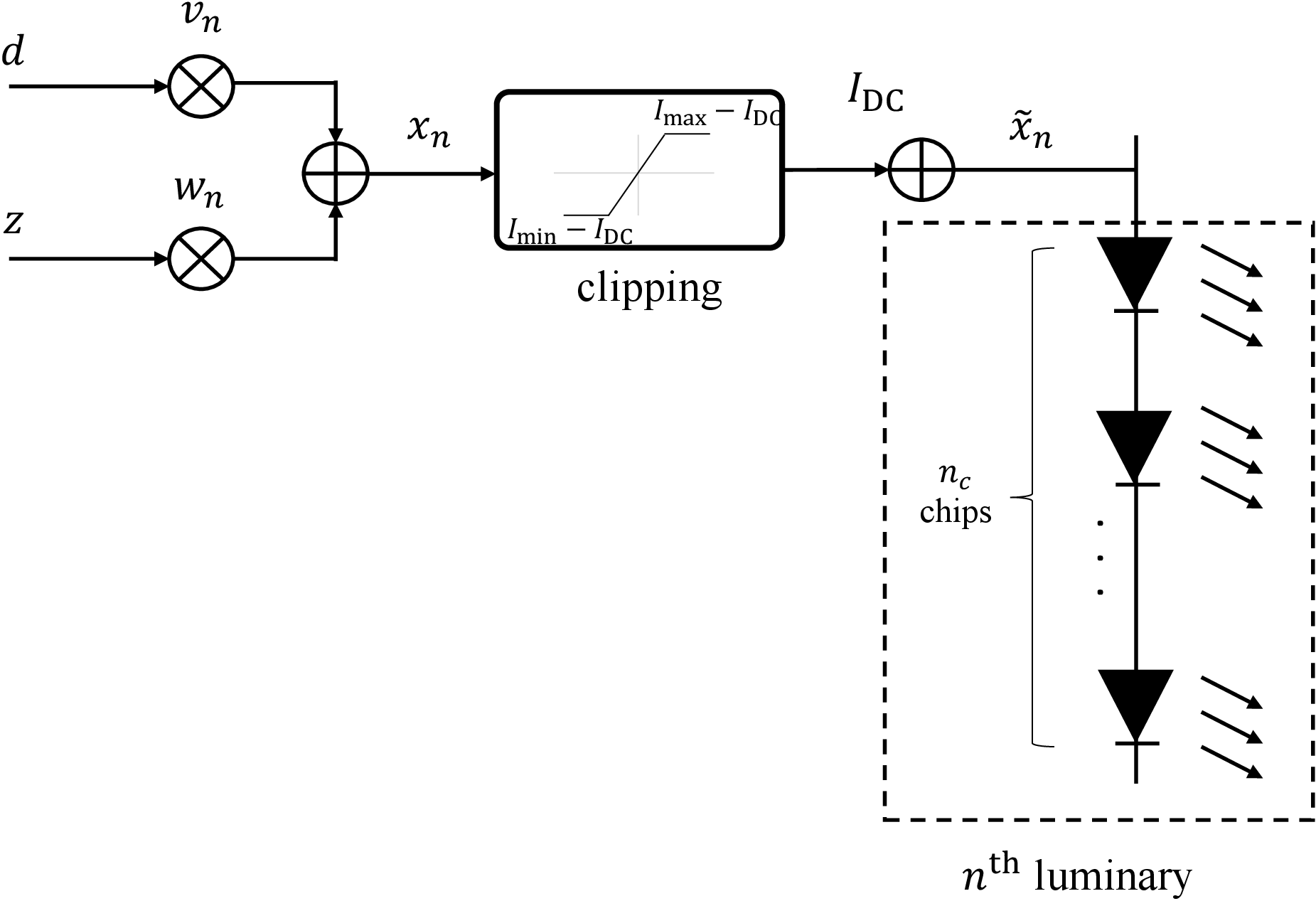}
    \caption{One-branch AN transmission scheme for non-amplitude constrained signals.}
    \label{Fig3}
\end{figure}
As illustrated in Fig.~\ref{Fig3}, a nonlinear clipping is applied to $x_n$ at a bottom level $I_{\text{min}}-I_{\text{DC}}$ and at a top level $I_{\text{max}}-I_{\text{DC}}$, where $I_{\text{DC}}$ is the DC-bias current which is added to the clipped signal. Thus, the resulting distorted signal $\Tilde{x}_n$ can be modeled by means of the Bussgang theorem as  
\begin{align}
    \Tilde{x}_n = R_n\left(v_nd + w_nz\right) + I_{\text{DC}} + \zeta_n,
    \label{clippedSignalConventional}
\end{align}
where $R_n \in [0,~1]$ is the attenuation factor and $\zeta_n$ is the clipping noise. An increased severity of clipping distortion is characterized by a small value of $R_n$ and a large variance of $\zeta_n$. If we denote $\alpha_n = \frac{I_{\text{min}} - I_{\text{DC}}}{\sqrt{v^2_n + w^2_n}}$ and $\beta_n = \frac{I_{\text{max}}- I_{\text{DC}}}{\sqrt{v^2_n + w^2_n}}$ then $R_n = Q(\alpha_n) - Q(\beta_n)$ \cite{Dimitrov2012}. The clipping noise $\zeta_n$ can be approximately modeled as a zero-mean Gaussian random variable whose variance is given by \cite{Dimitrov2012}
\begin{align}
    \sigma^2_{\text{clip}, n} =   &\Big(R_n+\alpha_n\phi(\alpha_n)-\beta_n\phi(\beta_n)  +\alpha_n^2(1-Q(\alpha_n))+\beta_n^2Q(\beta_n) \big. \nonumber \\ & \left. -\big(\phi(\alpha_n) - \phi(\beta_n) + (1-Q(\alpha_n))\alpha_n + Q(\beta_n)\beta_n\big)^2 \right. \!\!\! -  R_n^2\Big)  \times \left(v^2_n + w^2_n\right),
    \label{clipNoiseVariance}
\end{align}
where $\phi(t) = \frac{1}{2\pi}\exp\left(\frac{-t^2}{2}\right)$ and $Q(t) = \frac{1}{2\pi}\int_{t}^{\infty}\exp\left(\frac{-u^2}{2}\right)\text{d}u$.

Suppose that there are $n_c$ closely packed LED chips in each luminary. The instantaneous emitted optical power from the $n$-th LED luminary is then
\begin{align}
    P^{\text{(1-branch)}}_{n, \text{optical}} = n_c\eta\left(R_n\left(v_nd + w_nz\right) + I_{\text{DC}} + \zeta_n\right).
\end{align}
Since LED chips are close enough, it is reasonable to assume that the channel gains from them to a receiver are the same. The output electrical signal at Bob's PD after removing the DC current is then written by
\begin{align}
    r^{\text{(1-branch)}}_{\text{B}} = \gamma\eta n_c\left(\left(\mathbf{h}_{\text{B}}\odot\mathbf{R}\right)^T\mathbf{v}d  + \left(\mathbf{h}_{\text{B}}\odot\mathbf{R}\right)^T\mathbf{w}z + \mathbf{h}^T_{\text{B}}\pmb{\zeta}\right) + n^{\text{(1-branch)}}_{\text{B}},
    \label{user-received-signal}
\end{align}
where $\mathbf{R} = \begin{bmatrix}R_1 & R_2 & \cdots & R_{N_T}\end{bmatrix}^T$, $\mathbf{v} = \begin{bmatrix}v_1 & v_2 & \cdots & v_{N_T}\end{bmatrix}^T$, $\mathbf{w} = \begin{bmatrix}w_1 & w_2 & \cdots & w_{N_T}\end{bmatrix}^T$, and $\pmb{\zeta} = \begin{bmatrix}\zeta_1 & \zeta_2 & \cdots & \zeta_{N_T}\end{bmatrix}^T$. The receiver noise $n_{\text{B}}$ can be well approximated to be a Gaussian random variable, whose variance $\sigma^2_{\text{B}}$ is given by
\begin{align}
    \sigma^{2^{\text{(1-branch)}}}_{\text{B}} = B_{\text{mod}}\left(2e\gamma\overline{P}^{\text{(1-branch)}}_{r, \text{B}} + 4\pi e\gamma A_r\chi_{\text{amb}}\left(1 - \cos\Psi_c\right) + i^2_{\text{amp}}\right),
    \label{BobReceiverNoise}
\end{align}
where $e$ is the elementary charge, $B_{\text{mod}}$ is the signal bandwidth, $\chi_{\text{amb}}$ is the ambient light photocurrent, and $i^2_{\text{amp}}$ is the pre-amplifier noise current density. $\overline{P}^{\text{(1-branch)}}_{r, \text{B}} = n_c\eta \sum_{n=1}^{N_T}h_{n, \text{B}}\mathbb{E}[\Tilde{x}_n]$ is the average received optical power at Bob's receiver with $\mathbb{E}[\Tilde{x}_n]$ being
\begin{align}
     \mathbb{E}[\Tilde{x}_n] =  & \left(\phi(\alpha_n) - \phi(\beta_n) + \beta_nQ(\beta_n) - \alpha_nQ(\alpha_n)\right)\sqrt{v_n^2 + w_n^2}  + I_{\text{min}}.
     \label{meanClippedCurrent}
\end{align}
The SINR of Bob's received signal is then given by
\begin{align}
    \text{SINR}^{\text{(1-branch)}}_{\text{B}} = \frac{\left(n_c \left(\mathbf{h}_{\text{B}}\odot\mathbf{R}\right)^T\mathbf{v}\right)^2}{ \left(n_c\left(\mathbf{h}_{\text{B}}\odot\mathbf{R}\right)^T\mathbf{w}\right)^2 + \left(n_c\mathbf{h}^T_{\text{B}}\pmb{\sigma}_{\text{clip}}\right)^2 + {\sigma}^{2^{\text{(1-branch)}}}_{\text{B, norm}}},
    \label{BobSINR3}
\end{align}
where $\pmb{\sigma}_{\text{clip}} = \begin{bmatrix}\sigma_{\text{clip}, 1} & \sigma_{\text{clip}, 2}, & \hdots& \sigma_{\text{clip}, N_T}\end{bmatrix}^T$ and ${\sigma}^{2^{\text{(1-branch)}}}_{\text{B, norm}} = \frac{\sigma^{2^{\text{(1-branch)}}}_{\text{B}}}{\left(\gamma\eta\right)^2}$.
\subsection{Known $\mathbf{h}_{\rm{E}}$}
\label{KnownHe_AN}
Although it is often unrealistic to assume that $\mathbf{h}_{\text{E}}$ is known at the transmitter due to the fact that Eve tends to hide herself from being detected. Nonetheless, since the secrecy performance, in this case, represents an upper limit, it is worth studying AN designs under this assumption. 

Similar to \eqref{user-received-signal}, the received signal at Eve is written as
\begin{align}
    r^{\text{(1-branch)}}_{\text{E}}= \gamma\eta n_c\left(\left(\mathbf{h}_{\text{E}}\odot\mathbf{R}\right)^T\mathbf{v}d +  \left(\mathbf{h}_{\text{E}}\odot\mathbf{R}\right)^T\mathbf{w}z + \mathbf{h}^T_{\text{E}}\pmb{\zeta}\right) + n^{\text{(1-branch)}}_{\text{E}},
    \label{eve-received-signal}
\end{align}
where the noise variance $\sigma^2_{\text{E}}$ is given by
\begin{align}
    \sigma^{2^{\text{(1-branch)}}}_{\text{E}} = 2e\gamma\overline{P}^{\text{(1-branch)}}_{r, \text{E}}B_{\text{mod}} + 4\pi e\gamma A_r\chi_{\text{amb}}\left(1 - \cos\Psi_c\right)B_{\text{mod}} + i^2_{\text{amp}}B_{\text{mod}},
    \label{EveReceiverNoise}
\end{align}
with $\overline{P}^{\text{(1-branch)}}_{r, \text{E}} = n_c\eta\sum_{n = 1}^{N_T}h_{n, \text{E}}\mathbb{E}\left[\tilde{x}_n\right]$. The SINR of Eve's wire-taped signal is then given by
\begin{align}
    \text{SINR}^{\text{(1-branch)}}_{\text{E}} = \frac{\left(n_c \left(\mathbf{h}_{\text{E}}\odot\mathbf{R}\right)^T\mathbf{v}\right)^2}{ \left(n_c\left(\mathbf{h}_{\text{E}}\odot\mathbf{R}\right)^T\mathbf{w}\right)^2 + \left(n_c\mathbf{h}^T_{\text{E}}\pmb{\sigma}_{\text{clip}}\right)^2 + {\sigma}^{2^{\text{(1-branch)}}}_{\text{E, norm}}},
    \label{EveSINR3}
\end{align}
where $\sigma^{2^{\text{(1-branch)}}}_{\text{E, norm}} = \frac{\sigma^{2^{\text{(1-branch)}}}_{\text{E}}}{\left(\gamma\eta\right)^2}$.

Optimization problems in physical layer security often utilize the secrecy rate as the objective function\footnote{In optical systems, deriving the exact secrecy rate is often challenging. Therefore, an achievable secrecy rate is of practical interest in the system designs.}. In addition to the secrecy rate, the secrecy performance can also be evaluated indirectly through the pair of SINRs of Bob's and Eve's received signal, which is often easier to handle. Moreover, by taking the SINR of Eve's received signal into the design, one can precisely (or qualitatively) control the performance of Eve' channel, such as bit-error rate. In this work, we therefore study AN design problems using $\text{SINR}_{\text{B}}$ and $\text{SINR}_{\text{E}}$. 
Specifically, we aim at maximizing $\text{SINR}_{\text{B}}$ while trying to limit $\text{SINR}_{\text{E}}$ to a predefined threshold, say $\lambda$. Additionally, assume that the sum-power of the information-bearing and AN signals at the $n$-th luminary is constrained to a maximum allowable value of $P_n$, the AN design problem is given as follows
\begin{subequations}
\label{OptProb2}
    \begin{alignat}{2}
        &\underset{\mathbf{v}, \mathbf{w}}{\text{maximize}} & \hspace{2mm} & \text{SINR}^{\text{(1-branch)}}_{\text{B}} \label{obj2}\\
        &\text{subject to }  &  & \nonumber \\
        & & & \text{SINR}^{\text{(1-branch)}}_{\text{E}}  \leq \lambda, \label{constraint21}\\
        & & & \left[\mathbf{v}\right]^2_n + \left[\mathbf{w}\right]^2_n \leq P_n, ~~\forall n = 1, 2, ..., N_T \label{constraint22}.
    \end{alignat}
\end{subequations}
Note that, instead of \eqref{constraint22}, one can also apply a constraint on the sum-power of the information-bearing and AN signals over all luminaries, which is
\begin{align}
    \left\lVert\mathbf{v}\right\rVert^2 + \left\lVert\mathbf{w}\right\rVert^2 \leq P,
    \label{sumPowerConstraint}
\end{align}
where $P = \sum_{n = 1}^{N_T} P_n$. Compared with \eqref{sumPowerConstraint}, the advantage of using \eqref{constraint22} is two-fold. Firstly, it is clear that \eqref{constraint22} is a generalization of \eqref{sumPowerConstraint}. Secondly, using \eqref{constraint22} facilitates the handling of $\text{SINR}^{\text{(1-branch)}}_{\text{B}}$ and $\text{SINR}^{\text{(1-branch)}}_{\text{E}}$ as they are functions of $\left[\mathbf{v}\right]^2_n + \left[\mathbf{w}\right]^2_n$. Nonetheless, due to the complex non-linearity of $\mathbf{R}$, $\pmb{\sigma}_{\text{clip}}$, ${\sigma}^{2^{\text{(1-branch)}}}_{\text{B, norm}}$, and ${{\sigma}}^{2^{\text{(1-branch)}}}_{\text{E, norm}}$, it is extremely challenging (if not impossible) to optimally solve the above problem. Therefore, we examine a sub-optimal design by considering alternative simplified expressions for $\text{SINR}^{\text{(1-branch)}}_{\text{B}}$ and $\text{SINR}^{\text{(1-branch)}}_{\text{E}}$. Specifically, let use define $\widetilde{\mathbf{R}}$, $\widetilde{\pmb{\sigma}}_{\text{clip}}$, $\widetilde{{\sigma}}^{2^{\text{(1-branch)}}}_{\text{B, norm}}$, and $\widetilde{{{\sigma}}}^{2^{\text{(1-branch)}}}_{\text{E, norm}}$ be the values of $\mathbf{R}$, $\pmb{\sigma}_{\text{clip}}$, ${\sigma}^{2^{\text{(1-branch)}}}_{\text{B, norm}}$, and ${{{\sigma}}}^{2^{\text{(1-branch)}}}_{\text{E, norm}}$, respectively, at $\left[\mathbf{v}\right]^2_n +\left[\mathbf{w}\right]^2_n = P_n$  ($\forall n = 1, 2, \hdots, N_T$). Then, let $\widetilde{\text{SINR}}^{\text{(1-branch)}}_{\text{B}}$ and $\widetilde{\text{SINR}}^{\text{(1-branch)}}_{\text{E}}$ be the values of $\text{SINR}^{\text{(1-branch)}}_{\text{B}}$ and ${\text{SINR}}^{\text{(1-branch)}}_{\text{E}}$ at $\mathbf{R} = \widetilde{\mathbf{R}}$, $\pmb{\sigma}_{\text{clip}} = \widetilde{\pmb{\sigma}}_{\text{clip}}$, $\sigma^{2^{\text{(1-branch)}}}_{\text{B, norm}} = \widetilde{\sigma}^{2^{\text{(1-branch)}}}_{\text{B, norm}}$, and $\widetilde{{\sigma}}^{2^{\text{(1-branch)}}}_{\text{E, norm}} = {\sigma}^{2^{\text{(1-branch)}}}_{\text{E, norm}}$, respectively. We are now focusing on designing $\mathbf{v}$ and $\mathbf{w}$ based on $\widetilde{\text{SINR}}^{\text{(1-branch)}}_{\text{B}}$ and $\widetilde{\text{SINR}}^{\text{(1-branch)}}_{\text{E}}$. An intuitive reasoning for this sub-optimal design approach is that as long as \eqref{constraint21} is satisfied, $\left[\mathbf{v}\right]^2_n + \left[\mathbf{w}\right]^2_n$ tends to achieve its maximum allowable value so that \eqref{obj2} is maximized. Hence, when $\lambda$ is set not too stringent in relative to $P_n$'s, there is a high possibility that \eqref{constraint22} holds with equality at the optimal point. 

For the sake of analysis, let us define $\widetilde{\mathbf{h}}_{\text{B}} = \mathbf{h}_{\text{B}}\odot\widetilde{\mathbf{R}}$ and $\widetilde{\mathbf{h}}_{\text{E}} = {\mathbf{h}}_{\text{E}}\odot\widetilde{\mathbf{R}}$. The sub-optimal design problem is then given by
\begin{subequations}
\label{OptProb3}
    \begin{alignat}{2}
        &\underset{\mathbf{v}, \mathbf{w}}{\text{maximize}} & \hspace{2mm} & \widetilde{\text{SINR}}^{\text{(1-branch)}}_{\text{B}} \label{obj3}\\
        &\text{subject to }  &  & \nonumber \\
        & & & \widetilde{\text{SINR} }^{\text{(1-branch)}}_{\text{E}}  \leq \lambda, \label{constraint31}\\
        & & & \left[\mathbf{v}\right]^2_n + \left[\mathbf{w}\right]^2_n \leq P_n, ~~\forall n = 1, 2, ..., N_T \label{constraint32}.
    \end{alignat}
\end{subequations}
It is easy to see that there always exists $\mathbf{v}$ and $\mathbf{w}$ satisfying both \eqref{constraint31} and \eqref{constraint32} regardless of the values of $\lambda$ and $P_n$. Hence, \eqref{OptProb3} is always feasible. Also, it is seen that \eqref{OptProb3} is a fractional programming, which can often be solved using the Charnes-Cooper transformation \cite{Charnes1962}. Specifically, let us define
\begin{subequations}
\begin{flalign}
&&\mathbf{v} &= \frac{\overline{\mathbf{v}}}{t},\leftlabel{a} \text{and} & \mathbf{w} &= \frac{\overline{\mathbf{w}}}{t}, \label{b} &&
\end{flalign}
\end{subequations}
where $t > 0$. The problem in \eqref{OptProb3} can then be transformed into the following non-fractional programming
\begin{subequations}
\label{OptProb4}
    \begin{alignat}{2}
        &\underset{\overline{\mathbf{v}}, \overline{\mathbf{w}}, t}{\text{maximize}} & \hspace{0mm} & \left(n_{c}\widetilde{\mathbf{h}}^T_{\text{B}}\overline{\mathbf{v}}\right)^2 \label{obj4}\\
        &\text{subject to }  &  & \nonumber \\
        & & & \left(n_c\widetilde{\mathbf{h}}^T_{\text{B}}\overline{\mathbf{w}}\right)^2 + t^2\left(\left(n_c\mathbf{h}^T_{\text{B}}\widetilde{\pmb{\sigma}}_{\text{clip}}\right)^2 + \widetilde{\sigma}^{2^{\text{(1-branch)}}}_{\text{B, norm}}\right) = 1 \label{constraint41} \\
         & & & \frac{1}{\lambda}\left(n_c\widetilde{\mathbf{h}}^T_{\text{E}}\overline{\mathbf{v}}\right)^2 \leq \left(n_c\widetilde{\mathbf{h}}^T_{\text{E}}\overline{\mathbf{w}}\right)^2  + t^2\left(\left(n_c{\mathbf{h}}^T_{\text{E}}\widetilde{\pmb{\sigma}}_{\text{clip}}\right)^2+ \widetilde{{\sigma}}^{2^{\text{(1-branch)}}}_{\text{E, norm}}\right), \label{constraint42}\\
        & & & \left[\overline{\mathbf{v}}\right]^2_n + \left[\overline{\mathbf{w}}\right]^2_n \leq t^2P_n, ~~\forall n = 1, 2, \hdots, N_T. \label{constraint43}
    \end{alignat}
\end{subequations}
From \eqref{constraint41}, we get 
$
    t^2 = \frac{1-\left(n_c\widetilde{\mathbf{h}}^T_{\text{B}}\overline{\mathbf{w}}\right)^2}{\left(n_c\mathbf{h}^T_{\text{B}}\widetilde{\pmb{\sigma}}_{\text{clip}}\right)^2 + \widetilde{\sigma}^{2^{\text{(1-branch)}}}_{\text{B, norm}}}.
$ Replacing this into \eqref{constraint42} and \eqref{constraint43} results in 
\begin{subequations}
\label{OptProb5}
    \begin{alignat}{2}
        &\underset{\overline{\mathbf{v}}, \overline{\mathbf{w}}}{\text{maximize}}  \hspace{2mm}  \left(n_{c}\widetilde{\mathbf{h}}^T_{\text{B}}\overline{\mathbf{v}}\right)^2 & &\label{obj5}\\
        &\text{subject to }  &  & \nonumber \\
        & \frac{1}{\lambda}\left(n_c\widetilde{\mathbf{h}}^T_{\text{E}}\overline{\mathbf{v}}\right)^2 \leq \left(n_c\widetilde{\mathbf{h}}^T_{\text{E}}\overline{\mathbf{w}}\right)^2  + \frac{1-\left(n_c\widetilde{\mathbf{h}}^T_{\text{B}}\overline{\mathbf{w}}\right)^2}{\left(n_c\mathbf{h}^T_{\text{B}}\widetilde{\pmb{\sigma}}_{\text{clip}}\right)^2 + \widetilde{\sigma}^{2^{\text{(1-branch)}}}_{\text{B, norm}}}\left(\left(n_c{\mathbf{h}}^T_{\text{E}}\widetilde{\pmb{\sigma}}_{\text{clip}}\right)^2+ \widetilde{{\sigma}}^{2^{\text{(1-branch)}}}_{\text{E, norm}}\right), & & \label{constraint51}\\
        & \left[\overline{\mathbf{v}}\right]^2_n + \left[\overline{\mathbf{w}}\right]^2_n  \leq \frac{1-\left(n_c\widetilde{\mathbf{h}}^T_{\text{B}}\overline{\mathbf{w}}\right)^2}{\left(n_c\mathbf{h}^T_{\text{B}}\widetilde{\pmb{\sigma}}_{\text{clip}}\right)^2 + \widetilde{\sigma}^{2^{\text{(1-branch)}}}_{\text{B, norm}}}P_n, ~~\forall n = 1, \hdots, N_T. & &\label{constraint52}
    \end{alignat}
\end{subequations}
Obviously, the above optimization problem is not a convex due to the maximization of a convex function in \eqref{obj5} and the non-convexity of the constraint in \eqref{constraint51}. A well-known technique to solve \eqref{OptProb5} is CCP, which is based on an iterative procedure to find a local optimum \cite{yuille2003,lipp2016variations}. Specifically, at each iteration, the first-order Taylor approximation is utilized to approximately linearlize \eqref{obj5} and the first term in the right-hand side of \eqref{constraint51} as follows
\begin{align}
    \left(n_{c}\widetilde{\mathbf{h}}^T_{\text{B}}\overline{\mathbf{v}}^{(i)}\right)^2 \approx & \left(n_{c,1}\widetilde{\mathbf{h}}^T_{\text{B}}\overline{\mathbf{v}}^{(i-1)}\right)^2  2n^2_{c}\left[\overline{\mathbf{v}}^{(i-1)}\right]^T\widetilde{\mathbf{h}}_{\text{B}}\widetilde{\mathbf{h}}^T_{\text{B}}\left(\overline{\mathbf{v}}^{(i)} - \overline{\mathbf{v}}^{(i-1)} \right),
\end{align}
and
\begin{align}
    \left(n_c\widetilde{\mathbf{h}}^T_{\text{E}}\overline{\mathbf{w}}\right)^2
    \approx & n_c^2\Big(\left(\widetilde{\mathbf{h}}^T_{\text{E}}\overline{\mathbf{w}}^{(i-1)}\right)^2  +  2\left[\widetilde{\mathbf{h}}_{\text{E}}\widetilde{\mathbf{h}}^T_{\text{E}}\overline{\mathbf{w}}^{(i-1)}\right]^T\left(\overline{\mathbf{w}}^{(i)} - \overline{\mathbf{w}}^{(i-1)}\right)\Big),
\end{align}
where $\overline{\mathbf{v}}^{(i)}$ and $\overline{\mathbf{w}}^{(i)}$ are instances of $\overline{\mathbf{v}}$ and $\overline{\mathbf{w}}$ at the $i$-th iteration. Using the above approximations, the CCP involves solving the problem in \eqref{OptProb6}, which is convex. 
\begin{figure*}[h]
    \begin{subequations}
\label{OptProb6}
    \begin{alignat}{2}
        &\underset{\overline{\mathbf{v}}^{(i)}, \overline{\mathbf{w}}^{(i)}}{\text{maximize}} & \hspace{1mm} & \left(n_{c}\widetilde{\mathbf{h}}^T_{\text{B}}\overline{\mathbf{v}}^{(i-1)}\right)^2 + 2n^2_{c}\left[\overline{\mathbf{v}}^{(i-1)}\right]^T\widetilde{\mathbf{h}}_{\text{B}}\widetilde{\mathbf{h}}^T_{\text{B}}\left(\overline{\mathbf{v}}^{(i)} - \overline{\mathbf{v}}^{(i-1)} \right) \label{obj6}\\
        &\text{subject to }  &  & \nonumber \\
        & & & \frac{1}{\lambda}\left(n_{c}\widetilde{\mathbf{h}}^T_{\text{E}}\left[\overline{\mathbf{v}}^{(i)}\right]\right)^2 \leq n_c^2\left(\left(\widetilde{\mathbf{h}}^T_{\text{E}}\left[\overline{\mathbf{w}}^{(i-1)}\right]\right)^2  +  2\left[\widetilde{\mathbf{h}}_{\text{E}}\widetilde{\mathbf{h}}^T_{\text{E}}\overline{\mathbf{w}}^{(i-1)}\right]^T\left(\overline{\mathbf{w}}^{(i)} - \overline{\mathbf{w}}^{(i-1)}\right)\right) \nonumber \\  & & & \hspace{32mm} + \frac{1-\left(n_c\widetilde{\mathbf{h}}^T_{\text{B}}\overline{\mathbf{w}}\right)^2}{\left(n_c\mathbf{h}^T_{\text{B}}\widetilde{\pmb{\sigma}}_{\text{clip}}\right)^2 + \widetilde{\sigma}^{2^{\text{(1-branch)}}}_{\text{B, norm}}}\left(\left(n_c\mathbf{h}^T_{\text{E}}\widetilde{\pmb{\sigma}}_{\text{clip}}\right)^2+ \widetilde{{\sigma}}^{2^{\text{(1-branch)}}}_{\text{E, norm}}\right), \label{constraint61}\\
        & & & \left[\overline{\mathbf{v}}^{(i)}\right]_n^2 + \left[\overline{\mathbf{w}}^{(i)}\right]_n^2 \leq \frac{1-\left(n_c\widetilde{\mathbf{h}}^T_{\text{B}}\overline{\mathbf{w}}^{(i)}\right)^2}{\left(n_c\mathbf{h}^T_{\text{B}}\widetilde{\pmb{\sigma}}_{\text{clip}}\right)^2 + \widetilde{\sigma}^{2^{\text{(1-branch)}}}_{\text{B, norm}}}P_n, ~\forall n = 1, \hdots, N_T. \label{constraint62} 
    \end{alignat}
\end{subequations}
\end{figure*}
The detailed solving procedure is described in $\textbf{Algorithm 1}$. 
\begin{algorithm2e}[ht]
\SetAlgoLined 
\caption{CCP-type algorithm to solve \eqref{OptProb5}}
\label{alg.3}
Choose the maximum number of iterations $L$ and the error tolerance $\epsilon > 0$ for the algorithm. \\
Choose a feasible initial point $\left(\overline{\mathbf{v}}^{(0)}, \overline{\mathbf{w}}^{(0)}\right)$ to \eqref{OptProb6}. \\
Set $i \leftarrow 1$. \\
\While{convergence = \textbf{False} and $i \leq L$}{
Solve \eqref{OptProb6} using $\left(\overline{\mathbf{v}}^{(i-1)}, \overline{\mathbf{w}}^{(i-1)}\right)$ obtained from the previous iteration.\\
\eIf{$\frac{\left\lVert\overline{\mathbf{v}}^{(i)} - \overline{\mathbf{v}}^{(i-1)}\right\rVert}{\left\lVert\overline{\mathbf{v}}^{(i)}\right\rVert} \leq \epsilon$ and$\frac{\left\lVert\overline{\mathbf{w}}^{(i)} - \overline{\mathbf{w}}^{(i-1)}\right\rVert}{\left\lVert\overline{\mathbf{w}}^{(i)}\right\rVert} \leq \epsilon$}{
convergence = \textbf{True}. \\
$\overline{\mathbf{v}}^{*} \leftarrow \overline{\mathbf{v}}^{(i)}$. \\
$\overline{\mathbf{w}}^{*} \leftarrow \overline{\mathbf{w}}^{(i)}$. \\
}
{convergence = \textbf{False}.\\}
$i \leftarrow i + 1$. \\
}
Return the solution $\overline{\mathbf{v}}^{*}$ and $\overline{\mathbf{w}}^{*}$ then calculate  $\mathbf{v}^{*}$ and $\mathbf{w}^{*}$ from \eqref{a} and \eqref{b}.
\end{algorithm2e} 

The obtained sub-optimal precoders $\mathbf{v}^{*}$ and $\mathbf{w}^{*}$ are used to calculate the actual values of $\text{SINR}^{\text{(1-branch)}}_{\text{B}}$ and ${\text{SINR}}^{\text{(1-branch)}}_{\text{E}}$ given in \eqref{BobSINR3} and \eqref{avgEveSINR}, respectively. For comparisons, these values are denoted by $\text{SINR}^{*^{\text{(1-branch)}}}_{\text{B}}$ and ${\text{SINR}}^{*^{\text{(1-branch)}}}_{\text{E}}$ while the respective values resulted from solving \eqref{OptProb3} are denoted by $\widetilde{\text{SINR}}^{*^{\text{(1-branch)}}}_{\text{B}}$ and $\widetilde{{\text{SINR}}}^{*^{\text{(1-branch)}}}_{\text{E}}$. 

\subsection{Unknown $\mathbf{h}_{\rm{E}}$}
When $\mathbf{h}_{\text{E}}$ is unknown to the transmitters, it is not possible to use the instantaneous $\text{SINR}^{\text{(1-branch)}}_{\text{E}}$ for the optimal AN design as described in \eqref{OptProb2}. In this case, rather than using the instantaneous $\text{SINR}^{\text{(1-branch)}}_{\text{E}}$ for the AN design, one can consider the average SINR of $r_{\text{E}}$ over $\mathbf{h}_{\text{E}}$, which is written by
\begin{align}
    \overline{\text{SINR}}^{\text{(1-branch)}}_{\text{E}} = \mathbb{E}_{\mathbf{h}_{\text{E}}}\left[ \frac{\left(n_c\left(\mathbf{h}_{\text{E}}\odot\mathbf{R}\right)^T\mathbf{v}\right)^2}{\left(n_c\left(\mathbf{h}_{\text{E}}\odot\mathbf{R}\right)^T\mathbf{w}\right)^2 + \left(n_c\mathbf{h}^T_{\text{E}}\pmb{\sigma}_{\text{clip}}\right)^2 +  {\sigma}^{2^{\text{(1-branch)}}}_{\text{E, norm}}}\right].
    \label{avgEveSINR}
\end{align}
Nonetheless, the expression in \eqref{avgEveSINR} can not be straightforwardly transformed into a mathematically tractable form since $\text{SINR}^{\text{(1-branch)}}_{\text{E}}$ is a quadratic rational function of $\mathbf{h}_{\text{E}}$. Therefore, we employ the same approach proposed in \cite{Cho2019} by using the following expression as an estimation for $\overline{\text{SINR}}^{\text{(1-branch)}}_{\text{E}}$
\begin{align}
    \lambda^{\text{(1-branch)}}_{\text{E}} & =  \frac{\mathbb{E}_{\mathbf{h}_{\text{E}}}\left[\left(n_c\left(\mathbf{h}_{\text{E}}\odot\mathbf{R}\right)^T\mathbf{v}\right)^2\right]}{\mathbb{E}_{\mathbf{h}_{\text{E}}}\left[\left(n_c\left(\mathbf{h}_{\text{E}}\odot\mathbf{R}\right)^T\mathbf{w}\right)^2 + \left(n_c\mathbf{h}^T_{\text{E}}\pmb{\sigma}_{\text{clip}}\right)^2 +  {\sigma}^{2^{\text{(1-branch)}}}_{\text{E, norm}}\right]} \nonumber \\ & = \frac{n^2_c\mathbf{v}^T\left(\left(\mathbf{R}\mathbf{R}^T\right)\odot\overline{\mathbf{H}}_{\text{E}}\right)\mathbf{v}}{n^2_c\mathbf{w}^T\left(\left(\mathbf{R}\mathbf{R}^T\right)\odot\overline{\mathbf{H}}_{\text{E}}\right)\mathbf{w} + n^2_c\pmb{\sigma}^T_{\text{clip}}\overline{\mathbf{H}}_{\text{E}}\pmb{\sigma}_{\text{clip}} +  {\overline{\sigma}}^{2^{\text{(1-branch)}}}_{\text{E, norm}}},
    \label{lambdaE}
\end{align}
where $\overline{\mathbf{H}}_{\text{E}} = \mathbb{E}_{\mathbf{h}_{\text{E}}}\left[\mathbf{h}_{\text{E}}\mathbf{h}^T_{\text{E}}\right]$ and ${\overline{\sigma}}^{2^{\text{(1-branch)}}}_{\text{E, norm}} = \mathbb{E}_{\mathbf{h}_{\text{E}}}\left[{\sigma}^{2^{\text{(1-branch)}}}_{\text{E, norm}}\right]$, which are dependent on the room dimension and the luminary layout (e.g., number of luminaries and their positions).  In practice, it is reasonable to assume that these are fixed. To compute $\overline{\mathbf{H}}_{\text{E}}$ and ${\overline{\sigma}}^{2^{\text{(1-branch)}}}_{\text{E, norm}}$, let construct a 3D Cartesian coordinate system as illustrated in Fig.~\ref{Fig0} whose origin is the center of the floor. Then, $\overline{\sigma}^{2^{\text{(1-branch)}}}_{\text{E, norm}}$ is given by
\begin{align}
    {\overline{\sigma}}^{2^{\text{(1-branch)}}}_{\text{E, norm}}  = & 2e\gamma\mathbb{E}_{\mathbf{h}_{\text{E}}}\overline{P}^{\text{(1-branch)}}_{r, \text{E}}B_{\text{mod}}  + 4\pi e\gamma A_r\chi_{\text{amb}}\left(1 - \cos\Psi_c\right)B_{\text{mod}} + i^2_{\text{amp}}B_{\text{mod}} \nonumber \\
     = & \frac{2e\gamma n_c \eta B_{\text{mod}}}{d_{\text{L}}d_{\text{W}}}\int_{-\frac{d_{\text{L}}}{2}}^{\frac{d_{\text{L}}}{2}}\int_{-\frac{d_{\text{W}}}{2}}^{\frac{d_{\text{W}}}{2}}\sum_{n=1}^{N_T}h_{n, \text{E}}(x, y, d_{\text{R}})\mathbb{E}[\Tilde{x}_n]\text{d}x\text{d}y \nonumber \\
    & + 4\pi e\gamma A_r\chi_{\text{amb}}\left(1 - \cos\Psi_c\right)B_{\text{mod}} + i^2_{\text{amp}}B_{\text{mod}},
    \label{avgSigmaE}
\end{align}
where $h_{n, \text{E}}(x, y, d_{\text{R}})$ is the channel gain from the $n$-th LED luminary to the point with coordinate $(x, y, d_{\text{R}})$. The elements of $\overline{\mathbf{H}}_{\text{E}}$ can be calculated by
\begin{align}
    \left[\overline{\mathbf{H}}_{\text{E}}\right]_{m, n} = \frac{1}{d_{\text{L}}d_{\text{W}}}\int_{-\frac{d_{\text{L}}}{2}}^{\frac{d_{\text{L}}}{2}}\int_{-\frac{d_{\text{W}}}{2}}^{\frac{d_{\text{W}}}{2}}h_{m, \text{E}}(x, y, d_{\text{R}})h_{n, \text{E}}(x, y, d_{\text{R}})\text{d}x\text{d}y.
    \label{avgHE}
\end{align}

An average $\text{SINR}^{\text{(1-branch)}}_{\text{E}}$-based AN design problem is then given by
\begin{subequations}
\label{OptProb7}
    \begin{alignat}{2}
        &\underset{\mathbf{v}, \mathbf{w}}{\text{maximize}} & \hspace{2mm} & \text{SINR}^{\text{(1-branch)}}_{\text{B}} \label{obj7}\\
        &\text{subject to }  &  & \nonumber \\
        & & & \lambda^{\text{(1-branch)}}_{\text{E}}  \leq \lambda, \label{constraint71}\\
        & & & \left[\mathbf{v}\right]^2_n + \left[\mathbf{w}\right]^2_n \leq P_n, ~~\forall n = 1, 2, ..., N_T \label{constraint72}.
    \end{alignat}
\end{subequations}
First, define $\widetilde{\overline{\sigma}}^{2^{\text{(1-branch)}}}_{\text{E, norm}}$ as the value of $\overline{\sigma}^{2^{\text{(1-branch)}}}_{\text{E, norm}}$ at $\left[\mathbf{v}\right]^2_n  + \left[\mathbf{w}\right]^2_n = P_n$ ($\forall n = 1, 2, \hdots, N_T$). Then, let  
 $\widetilde{\lambda}^{\text{(1-branch)}}_{\text{E}}$ be the value of $\lambda_{\text{E}}$ at $\mathbf{R} = \widetilde{\mathbf{R}}$, $\pmb{\sigma}_{\text{clip}} = \widetilde{\pmb{\sigma}}_{\text{clip}}$, and $\overline{\sigma}^{2^{\text{(1-branch)}}}_{\text{E, norm}} = \widetilde{\overline{\sigma}}^{2^{\text{(1-branch)}}}_{\text{E, norm}}$. Similar to the case of known $\mathbf{h}_{\text{E}}$, we consider an alternative optimization problem to \eqref{OptProb7} as follows 
 \begin{subequations}
\label{OptProb8}
    \begin{alignat}{2}
        &\underset{\mathbf{v}, \mathbf{w}}{\text{maximize}} & \hspace{2mm} & \widetilde{\text{SINR}}^{\text{(1-branch)}}_{\text{B}} \label{obj8}\\
        &\text{subject to }  &  & \nonumber \\
        & & & \widetilde{\lambda}^{\text{(1-branch)}}_{\text{E}}  \leq \lambda, \label{constraint81}\\
        & & & \left[\mathbf{v}\right]^2_n + \left[\mathbf{w}\right]^2_n \leq P_n, ~~\forall n = 1, 2, ..., N_T \label{constraint82}.
    \end{alignat}
\end{subequations}
Based on the sample approach presented for handling \eqref{OptProb3} and denoting  $\widetilde{\overline{\mathbf{H}}}_{\text{E}} = \left(\mathbf{R}\mathbf{R}^T\right)\odot\overline{\mathbf{H}}_{\text{E}}$, sub-optimal solutions to \eqref{OptProb8} can be found via applying \textbf{Algorithm 1} for the surrogate problem in \eqref{OptProb8-1}, which is on top of the next page.
\begin{figure*}[ht]
    \begin{subequations}
\label{OptProb8-1}
    \begin{alignat}{2}
        &\underset{\overline{\mathbf{v}}^{(i)}, \overline{\mathbf{w}}^{(i)}}{\text{maximize}} & \hspace{1mm} & \left(n_{c}\widetilde{\mathbf{h}}^T_{\text{B}}\overline{\mathbf{v}}^{(i-1)}\right)^2 + 2n^2_{c}\left[\overline{\mathbf{v}}^{(i-1)}\right]^T\widetilde{\mathbf{h}}_{\text{B}}\widetilde{\mathbf{h}}^T_{\text{B}}\left(\overline{\mathbf{v}}^{(i)} - \overline{\mathbf{v}}^{(i-1)} \right) \label{obj8-1}\\
        &\text{subject to }  &  & \nonumber \\
        & & & \frac{1}{\lambda}n^2_{c}\left[\overline{\mathbf{v}}^{(i)}\right]^T\widetilde{\overline{\mathbf{H}}}_{\text{E}}\overline{\mathbf{v}}^{(i)} \leq n_c^2\left(\left[\overline{\mathbf{w}}^{(i-1)}\right]^T\widetilde{\overline{\mathbf{H}}}_{\text{E}}\overline{\mathbf{w}}^{(i-1)}  +  2\left[\widetilde{\overline{\mathbf{H}}}_{\text{E}}\overline{\mathbf{w}}^{(i-1)}\right]^T\left(\overline{\mathbf{w}}^{(i)} - \overline{\mathbf{w}}^{(i-1)}\right)\right) \nonumber \\  & & & \hspace{32mm} + \frac{1-\left(n_c\widetilde{\mathbf{h}}^T_{\text{B}}\overline{\mathbf{w}}\right)^2}{\left(n_c\mathbf{h}^T_{\text{B}}\widetilde{\pmb{\sigma}}_{\text{clip}}\right)^2 + \widetilde{\sigma}^{2^{\text{(1-branch)}}}_{\text{B, norm}}}\left(n^2_c\widetilde{\pmb{\sigma}}^T_{\text{clip}}{\overline{\mathbf{H}}}_{\text{E}}\widetilde{\pmb{\sigma}}_{\text{clip}}+ \widetilde{\overline{\sigma}}^{2^{\text{(1-branch)}}}_{\text{E, norm}}\right), \label{constraint8-11}\\
        & & & \left[\overline{\mathbf{v}}^{(i)}\right]_n^2 + \left[\overline{\mathbf{w}}^{(i)}\right]_n^2 \leq \frac{1-\left(n_c\widetilde{\mathbf{h}}^T_{\text{B}}\overline{\mathbf{w}}^{(i)}\right)^2}{\left(n_c\mathbf{h}^T_{\text{B}}\widetilde{\pmb{\sigma}}_{\text{clip}}\right)^2 + \widetilde{\sigma}^{2^{\text{(1-branch)}}}_{\text{B, norm}}}P_n, ~\forall n = 1, \hdots, N_T. \label{constraint8-12} 
    \end{alignat}
\end{subequations}
\end{figure*}
The obtained precoder $\mathbf{v}^*$ and $\mathbf{w}^*$ in this case are then used to compute $\text{SINR}^{*^{\text{(1-branch)}}}_{\text{B}}$, $\overline{\text{SINR}}^{*^{\text{(1-branch)}}}_{\text{E}}$, and $\lambda^{*^{\text{(1-branch)}}}_{\text{E}}$ given in \eqref{BobSINR3}, \eqref{avgEveSINR}, and \eqref{lambdaE}, respectively. Note that, for each pair of $\mathbf{v}^*$ and $\mathbf{w}^*$, $\overline{\text{SINR}}^{*^{\text{(1-branch)}}}_{\text{E}}$ is numerically evaluated through averaging 5000 different realizations of $\mathbf{h}_{\text{E}}$. Furthermore, the resulting $\widetilde{\lambda}^{\text{(1-branch)}}_{\text{E}}$ from solving \eqref{OptProb8} is denoted as $\widetilde{\lambda}^{*^{\text{(1-branch)}}}_{\text{E}}$. 
\section{Proposed Two-Branch AN transmission scheme}
\label{sec:proposedAN}
\subsection{Impact of Clipping Distortion}
\begin{figure}[ht]
    \centering
    \includegraphics[width = 0.72\textwidth, height = 7.0cm]{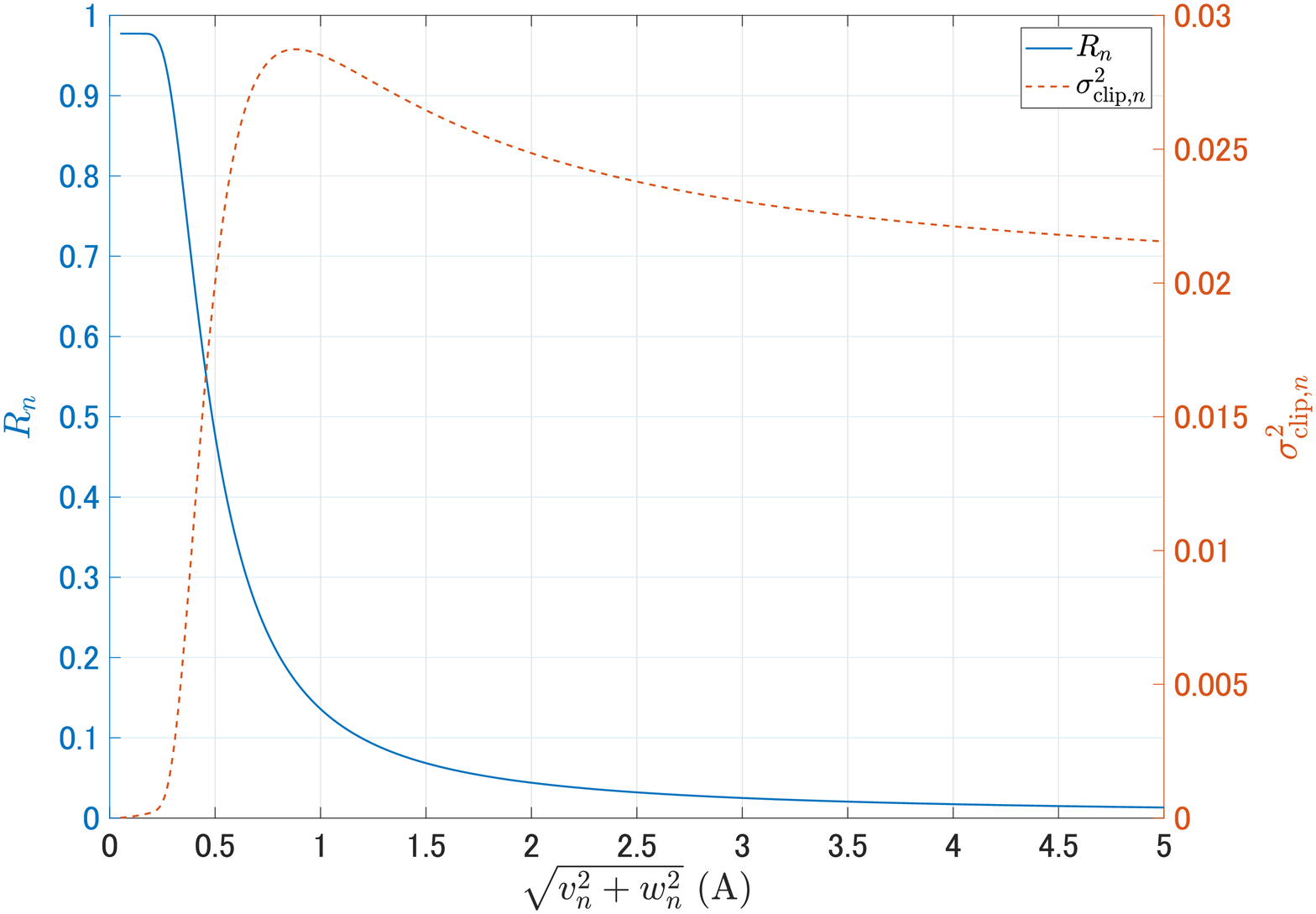}
    \caption{Effect of clipping distortion ($I_{\text{min}} = 0$ A, $I_{\text{max}} = 1$ A).}
    \label{Fig4}
\end{figure}
Given a fixed LED linear range $[I_{\text{min}},~I_{\text{max}}]$, it is known that clipping distortion can severely degrade the signal quality, especially in the case of large variations of the signal amplitude. This negative impact is represented in terms of the attenuation factor $R_n$ and the clipping noise power as shown in Fig.~\ref{Fig4} where suppose that $I_{\text{min}}= 0$ A and $I_{\text{max}} = 1$ A  and $\sqrt{v_n^2 + w_n^2}$ is the standard deviation of the combined information-bearing and AN signals. Assuming DCO-OFDM is employed to generate the information-bearing signal and AN, a widely used setting for the DC-biased current is $I_{\text{DC}} = 2\sqrt{v_n^2 + w_n^2}$,  which is equivalent to 7 dB bias \cite{Armstrong2008}. It is clear that as the standard deviation of the combined signal increases, the attenuation factor decreases rapidly, which indicates an increased severity of the clipping distortion since $R_n$ has a multiplicative impact on the SINR of the received signal as seen from \eqref{BobSINR3}. It is also noticed that the clipping noise variance increases quickly as $\sqrt{v_n^2 + w_n^2}$ increases until a certain value (e.g., about $0.9$ A in this case). However, it then starts decreasing due to the small value of $R_n$, which becomes the dominant term in reducing the clipping noise variance as seen from \eqref{clipNoiseVariance}. 
\begin{figure*}[ht]
    \centering
    \begin{subfigure}[b]{\textwidth}
        \centering
        \includegraphics[width = .85\textwidth, height = 4.8cm]{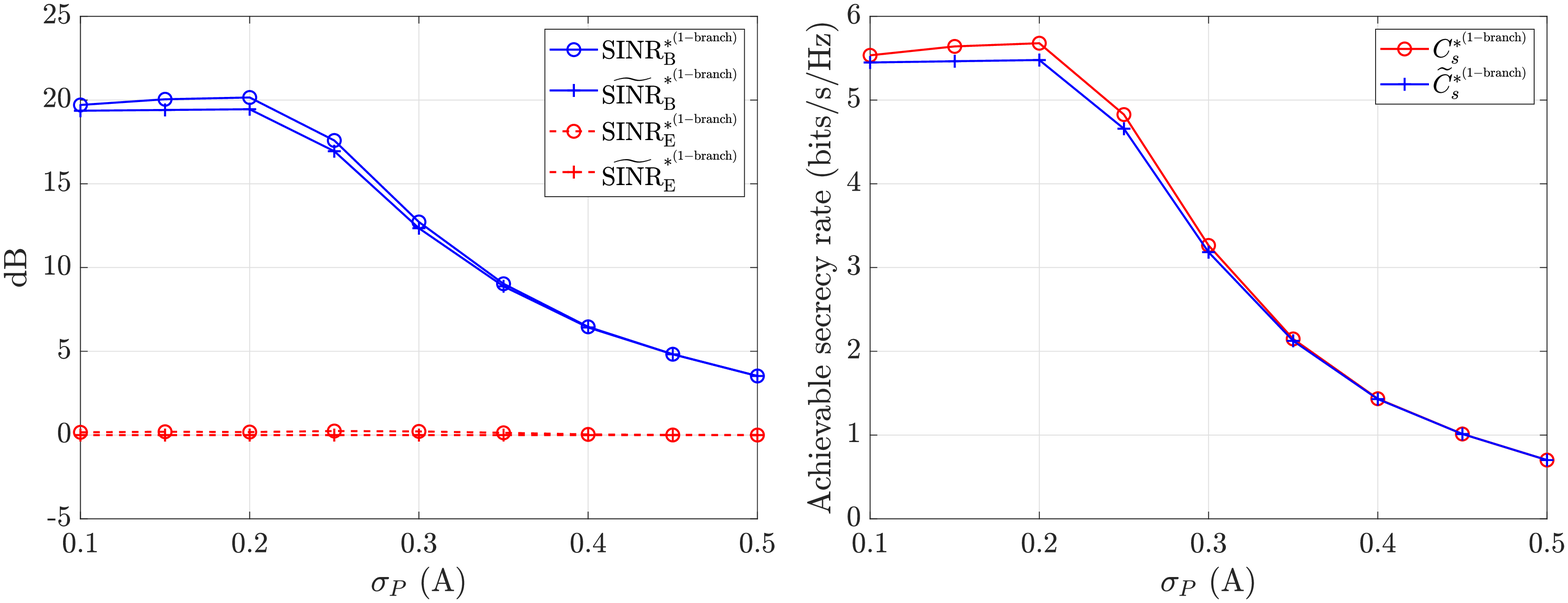}
        \caption{Known $\mathbf{h}_{\text{E}}$.}
        \label{KnownHe1}
    \end{subfigure}
    \begin{subfigure}[b]{\textwidth}
        \centering
        \includegraphics[width = .85\textwidth, height = 4.8cm]{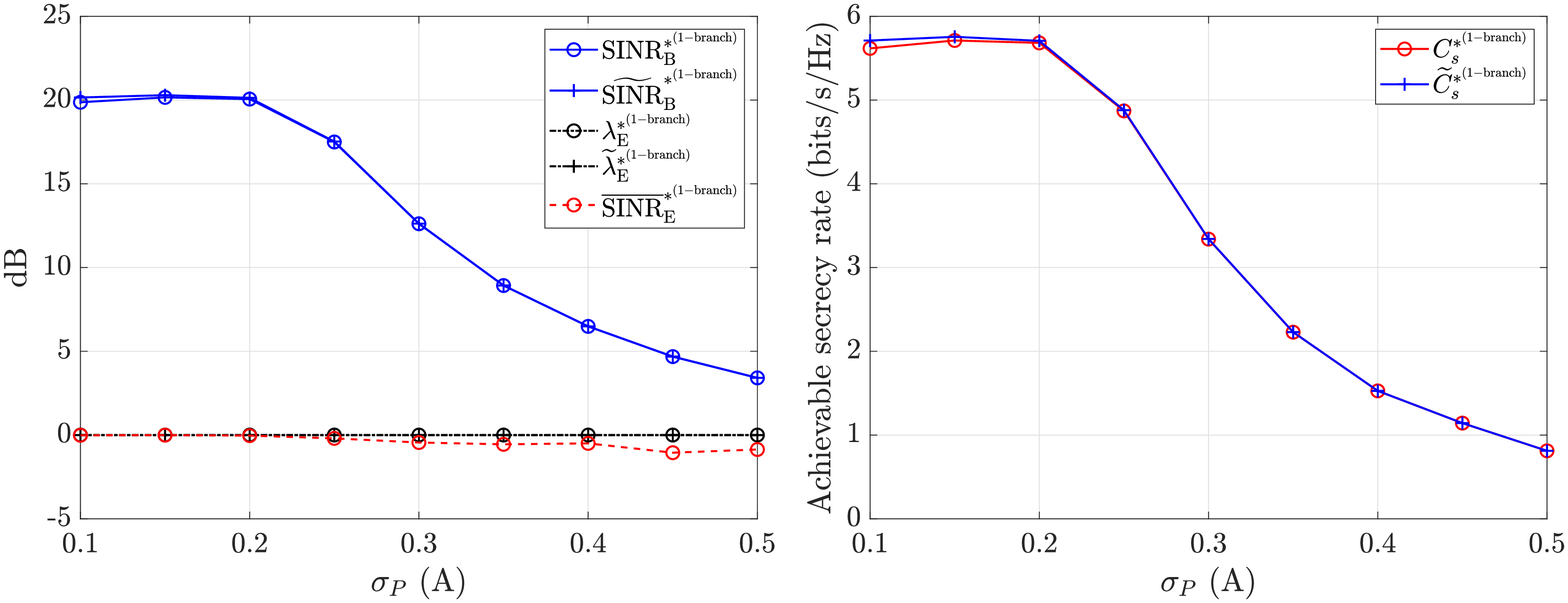}
        \caption{Unknown $\mathbf{h}_{\text{E}}$.}
        \label{UnknownHe1}
    \end{subfigure}
    \caption{Secrecy performances versus $\sigma_{P}$ with $\lambda = 0$ dB.}
\end{figure*}

To directly illustrate the impact of the clipping distortion, Fig.~\ref{KnownHe1} and Fig.~\ref{UnknownHe1} show the secrecy performances in case of known $\mathbf{h}_{\text{E}}$ and unknown $\mathbf{h}_{\text{E}}$ by solving problems  \eqref{OptProb3} and \eqref{OptProb8}, respectively. Here, $\lambda = 0$ dB is set and assume that the same maximum allowable power $P$ is allocated for the combined information-bearing and AN signals in each luminary (i.e., $P_n$'s = $P$) and denote $\sigma_P = \sqrt{P}$. Other simulation parameters are given in Sec.~\ref{sec:numerical-discussions} Table I. Simulation results are obtained through averaging over 5000 random realizations of $\mathbf{h}_{\text{B}}$ and $\mathbf{h}_{\text{E}}$. Firstly, it is observed in the left-hand side figures that, while Eve's SINRs stay almost unchanged, Bob's SINRs decrease greatly as $\sigma_P$ increases (especially beyond 0.2 A). For the sake of conciseness, a detailed explanation for this is given in Sec.~\ref{sec:numerical-discussions} of the paper. Since the information-bearing and AN signals are approximated to be Gaussian, the achievable secrecy rates corresponding to the pairs $\left(\text{SINR}^{*^{\text{(1-branch)}}}_{\text{B}},~\text{SINR}^{*^{\text{(1-branch)}}}_{\text{E}}\right)$ and $\left(\widetilde{\text{SINR}}^{*^{\text{(1-branch)}}}_{\text{B}}, ~\widetilde{\text{SINR}}^{*^{\text{(1-branch)}}}_{\text{E}}\right)$ in the case of known $\mathbf{h}_{\text{E}}$ are given by $C^{*^{\text{(1-branch)}}}_s = \log_2\left(1 + \text{SINR}^{*^{\text{(1-branch)}}}_{\text{B}}\right) - \log_2\left(1 + {\text{SINR}}^{*^{\text{(1-branch)}}}_{\text{E}}\right)$ and $\widetilde{C}^*_s = \log_2\left(1 + \widetilde{\text{SINR}}^{*^{\text{(1-branch)}}}_{\text{B}}\right) - \log_2\left(1 + \widetilde{{\text{SINR}}}^{*^{\text{(1-branch)}}}_{\text{E}}\right)$, respectively. Similarly, in the case of unknown $\mathbf{h}_{\text{E}}$, $C^{*^{\text{(1-branch)}}}_s$ and $\widetilde{C}^{*^{\text{(1-branch)}}}_s$ are defined by $C^{*^{\text{(1-branch)}}}_s = \log_2\left(1 + \text{SINR}^{*^{\text{(1-branch)}}}_{\text{B}}\right) - \log_2\left(1 + \overline{\text{SINR}}^{*^{\text{(1-branch)}}}_{\text{E}}\right)$ and $\widetilde{C}^{*^{\text{(1-branch)}}}_s = \log_2\left(1 + \widetilde{\text{SINR}}^{*^{\text{(1-branch)}}}_{\text{B}}\right) - \log_2\left(1 + \overline{{\text{SINR}}}^{*^{\text{(1-branch)}}}_{\text{E}}\right)$. Then, significant reductions in the achievable secrecy rate due to clipping distortion are clearly displayed in the right-hand side figures. 
\subsection{Proposed Transmission Scheme}
\begin{figure*}[ht]
    \centering
    \includegraphics[width = 0.95\textwidth, height = 5.0cm]{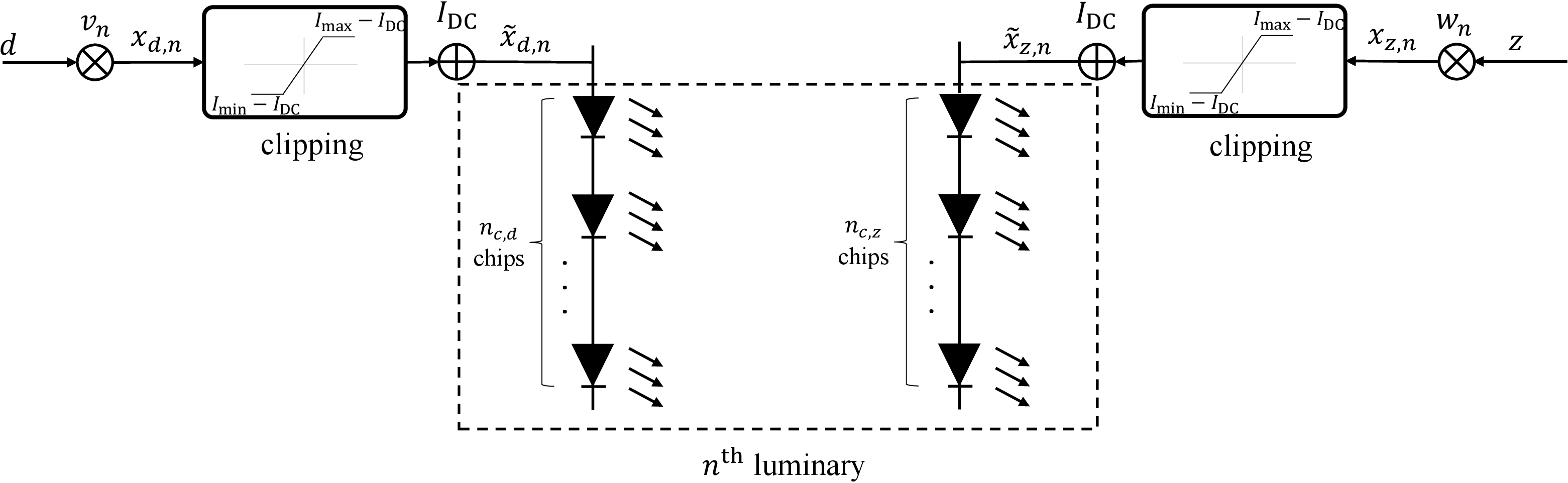}
    \caption{Proposed two-branch AN transmission scheme for non-amplitude constrained signals.}
    \label{Fig5}
\end{figure*}
In practice, a certain LED has a fixed linear range. Hence, to alleviate this negative impact of clipping distortion on the secrecy performance, a straightforward approach is to reduce the signal power flowing through each LED chip. In the one-branch AN scheme, each LED chip is driven by the combined information-bearing and AN signals that potentially results in a high clipping. Since the clipping severity is proportional to the signal power, it can be reduced if each LED chip is driven by either information-bearing or AN signal. Therefore, in our proposed AN transmission scheme illustrated in Fig. \ref{Fig5}, LED chips in each luminary are divided into two groups with separate circuits. Then, the first group is driven by the information-bearing signal, while the second group is driven by the AN. At the $n$-th luminary, let $x_{d, n}$ be the combined DC current, and information-bearing signal, and $x_{z, n}$ be the combined DC current and AN. According to the Bussgang theorem, the clipped signals of $x_{d, n}$ and $x_{z, n}$ are given by 
\begin{align}
    \Tilde{x}_{d, n} = R_{d, n}v_nd + I_{\text{DC}} + \zeta_{d, n},
    \label{clippedSignal}
\end{align}
and 
\begin{align}
    \Tilde{x}_{z, n} = R_{z, n}w_nz + I_{\text{DC}} + \zeta_{z, n},
\end{align}
where $R_{d, n}$ and $R_{z, n}$ are the attenuation factors while $\zeta_{d, n}$ and $\zeta_{z, n}$ are the clipping noises. Denote $\alpha_{d, n} = \frac{I_{\text{min}} - I_{\text{DC}}}{\left|v_n\right|}$, $\alpha_{z, n} = \frac{I_{\text{min}} - I_{\text{DC}}}{\left|w_n\right|}$, $\beta_{d, n} = \frac{I_{\text{max}} - I_{\text{DC}}}{\left|v_n\right|}$, and
$\beta{z, n} = \frac{I_{\text{max}} - I_{\text{DC}}}{\left|w_n\right|}$. Then, we have $R_{d, n} = Q(\alpha_{d, n}) - Q(\beta_{d, n})$ and $R_{z, n} = Q(\alpha_{z, n}) - Q(\beta_{z, n})$. The variances of $\zeta_{d, n}$ and $\zeta_{z, n}$ are also given by
\begin{align}
    \sigma^2_{\text{clip}, d, n} =  &\Big(R_{d,n}+\alpha_n\phi(\alpha_{d, n})-\beta_{d, n}\phi(\beta_{d, n})  \left. +\alpha_{d, n}^2(1-Q(\alpha_{d, n}))  +\beta_{d, n}^2Q(\beta_{d, n}) \right. \nonumber \\ & \left. -  \big(\phi(\alpha_{d,n}) - \phi(\beta_{d,n}) + (1-Q(\alpha_{d, n}))\alpha_{d, n} + Q(\beta_{d, n})\beta_{d, n}\big)^2 \right.  - R_{d, n}^2\Big)  v^2_n,
    \label{Bob-clipped-received-signal}
\end{align}
and
\begin{align}
    \sigma^2_{\text{clip}, z, n} =  &\Big(R_{z,n}+\alpha_n\phi(\alpha_{z, n})-\beta_{z, n}\phi(\beta_{z, n}) \big. \left. +\alpha_{z, n}^2(1-Q(\alpha_{z, n}))  +\beta_{z, n}^2Q(\beta_{z, n}) \right. \nonumber \\ & \left. - \big(\phi(\alpha_{z,n}) - \phi(\beta_{z,n}) + (1-Q(\alpha_{z, n}))\alpha_{z, n} + Q(\beta_{z, n})\beta_{z, n}\big)^2 \right.   - R_{z, n}^2\Big)w^2_n,
\end{align}
respectively. 

Assume that there are $n_{c ,d}$ and $n_{c, z}$ chips in the first and second groups, respectively. The average emitted optical power of each luminary is then 
\begin{align}
    \overline{P}^{\text{(2-branch)}}_{n, \text{optical}} = &  \eta\left(n_{c,d}\mathbb{E}[\Tilde{x}_{d,n}] + n_{c, z}\mathbb{E}[\Tilde{x}_{z, n}]\right) \nonumber \\  = &
    \eta\Big(n_{c, d}\left(\left(\phi(\alpha_{d, n}) - \phi(\beta_{d,n}) + \beta_{d, n}Q(\beta_{d,n}) - \alpha_{d, n}Q(\alpha_{d, n})\right)|v_n| + I_{\text{min}}\right) \big. \nonumber \\ & + n_{c, z}\left(\left(\phi(\alpha_{z, n}) - \phi(\beta_{z,n}) + \beta_{z, n}Q(\beta_{z,n}) - \alpha_{z, n}Q(\alpha_{z, n})\right)|w_n|  + I_{\text{min}}\right)\Big).
    \label{meanEmitOpticalPowerProposed}
\end{align}
The received signals at Bob and Eve are then given by
\begin{align}
    r^{\text{(2-branch)}}_{\text{B}} = & \gamma\eta\Bigg(n_{c,d}\bigg(\left(\mathbf{h}_{\text{B}}\odot\mathbf{R}_{d}\right)^T{\mathbf{v}}d + \mathbf{h}^T_{\text{B}}\pmb{\zeta}_{d}\bigg)  + n_{c, z}\bigg(\left(\mathbf{h}_{\text{B}}\odot\mathbf{R}_{z}\right)^T{\mathbf{w}}z+\mathbf{h}^T_{\text{B}}\pmb{\zeta}_{z}\bigg)\Bigg)  + n^{\text{(2-branch)}}_{\text{B}}
\end{align}
and
\begin{align}
    r^{\text{(2-branch)}}_{\text{E}} = & \gamma\eta\Bigg(n_{c,d}\bigg(\left(\mathbf{h}_{\text{E}}\odot\mathbf{R}_{d}\right)^T{\mathbf{v}}d + \mathbf{h}^T_{\text{E}}\pmb{\zeta}_{d}\bigg)   + n_{c, z}\bigg(\left(\mathbf{h}_{\text{E}}\odot\mathbf{R}_{z}\right)^T{\mathbf{w}}z+\mathbf{h}^T_{\text{E}}\pmb{\zeta}_{z}\bigg)\Bigg)  + n^{\text{(2-branch)}}_{\text{E}},
\end{align}
respectively, where $\mathbf{R}_d = \begin{bmatrix}R_{d, 1} & R_{d, 2} \cdots & R_{d, N_T}\end{bmatrix}^T$, $\pmb{\zeta}_d = \begin{bmatrix}\zeta_{d, 1} & \zeta_{d, 2} & \cdots & \zeta_{d, N_T}\end{bmatrix}^T$, and $\pmb{\zeta}_z = \begin{bmatrix}\zeta_{z,1} & \zeta_{z, 2} & \cdots & \zeta_{z, N_T}\end{bmatrix}^T$.  

In this proposed scheme, one can flexibly adjust power allocations for the information-bearing and AN signals by changing $n_{c, d}$ and $n_{c, z}$, which in turn influences the secrecy performance. Optimal choices for ($n_{c, d}$, $n_{c, z}$) under different parameter settings will be numerically discussed in the next section. From \eqref{Bob-clipped-received-signal}, the SINR of Bob's received signal is written by
\begin{align}
    \text{SINR}^{\text{(2-branch)}}_{\text{B}} = \frac{\left(n_{c,d}\left(\mathbf{h}_{\text{B}}\odot\mathbf{R}_{d}\right)^T{\mathbf{v}}\right)^2}{ \left(n_{c,d}\mathbf{h}^T_{\text{B}}\pmb{\sigma}_{\text{clip}, d}\right)^2 + n_{c,z}^2\left(\left(\left(\mathbf{h}_{\text{B}}\odot\mathbf{R}_{z}\right)^T{\mathbf{w}}\right)^2 + \left(\mathbf{h}^T_{\text{B}}\pmb{\sigma}_{\text{clip}, z}\right)^2\right) + {\sigma}^{2^{\text{(2-branch)}}}_{\text{B},\text{norm}}},
    \label{BobSINR4}
\end{align}
where $\pmb{\sigma}_{\text{clip}, d} = \begin{bmatrix}\sigma_{\text{clip}, d, 1} & \sigma_{\text{clip}, d, 2} & \hdots& \sigma_{\text{clip}, d, N_T}\end{bmatrix}^T$ and $\pmb{\sigma}_{\text{clip}, z} = \begin{bmatrix}\sigma_{\text{clip}, z, 1} & \sigma_{\text{clip}, z, 2} & \hdots& \sigma_{\text{clip}, z, N_T}\end{bmatrix}^T$. Given the average emitted optical power of each luminary in \eqref{meanEmitOpticalPowerProposed}, the noise power $\sigma^{2^{\text{(2-branch)}}}_{\text{B}, \text{norm}}$ is given by
\begin{align}
    \sigma^{2^{\text{(2-branch)}}}_{\text{B}, \text{norm}} = \frac{B_{\text{mod}}}{\left(\gamma\eta\right)^2}\left(2e\gamma\sum_{n = 1}^{N_T}h_{n, \text{B}}\overline{P}^{\text{(2-branch)}}_{n, \text{optical}} + 4\pi e\gamma A_r\chi_{\text{amb}}\left(1 - \cos\Psi_c\right) + i^2_{\text{amp}}\right).
\end{align}
In the following, AN designs for the proposed scheme under known and unknown $\mathbf{h}_{\text{E}}$ scenarios are then described. 
\subsection{Known $\mathbf{h}_{\rm{E}}$}
When $\mathbf{h}_{\text{E}}$ is known at the transmitter, the instantaneous SINR of Eve's eavesdropped signal is employed for the AN design, which is given by
\begin{align}
\text{SINR}^{\text{(2-branch)}}_{\text{E}} =\frac{\left(n_{c,d}\left(\mathbf{h}_{\text{E}}\odot\mathbf{R}_{d}\right)^T{\mathbf{v}}\right)^2}{ \left(n_{c,d}\mathbf{h}^T_{\text{E}}\pmb{\sigma}_{\text{clip}, d}\right)^2 + n_{c,z}^2\left(\left(\left(\mathbf{h}_{\text{E}}\odot\mathbf{R}_{z}\right)^T{\mathbf{w}}\right)^2 + \left(\mathbf{h}^T_{\text{E}}\pmb{\sigma}_{\text{clip}, z}\right)^2\right) + {\sigma}^{2^{\text{(2-branch)}}}_{\text{E, norm}}},
\end{align}
where 
\begin{align}
     \sigma^{2^{\text{(2-branch)}}}_{\text{E}, \text{norm}} = \frac{B_{\text{mod}}}{\left(\gamma\eta\right)^2}\left(2e\gamma\sum_{n = 1}^{N_T}h_{n, \text{E}}\overline{P}^{\text{(2-branch)}}_{n, \text{optical}} + 4\pi e\gamma A_r\chi_{\text{amb}}\left(1 - \cos\Psi_c\right) + i^2_{\text{amp}}\right).
\end{align}
The AN design problem in this case is then given by
\begin{subequations}
\label{OptProb9}
    \begin{alignat}{2}
        &\underset{\mathbf{v}, \mathbf{w}}{\text{maximize}} & \hspace{2mm} & \text{SINR}^{\text{(2-branch)}}_{\text{B}} \label{obj9}\\
        &\text{subject to }  &  & \nonumber \\
        & & & \text{SINR}^{\text{(2-branch)}}_{\text{E}}  \leq \lambda, \label{constraint91}\\
        & & & \left[\mathbf{v}\right]_n^2 + \left[\mathbf{w}\right]_n^2 \leq P_n,~~ \forall n=1, 2, ..., N_T. \label{constraint92}
    \end{alignat}
\end{subequations}
Similar to the one-branch AN transmission scheme, we are interested in a suboptimal solution to \eqref{OptProb9} by considering simpler alternative expressions for $\text{SINR}^{\text{(2-branch)}}_{\text{B}}$ and $\text{SINR}^{\text{(2-branch)}}_{\text{E}}$. However, it can be seen that $\mathbf{R}_d$, $\mathbf{R}_z$, $\pmb{\sigma}_{\text{clip}, d}$, $\pmb{\sigma}_{\text{clip}, z}$, $\sigma^{2^{\text{(2-branch)}}}_{\text{B, norm}}$, and  $\sigma^{2^{\text{(2-branch)}}}_{\text{E, norm}}$ are not functions of $\left[\mathbf{v}\right]^2_n + \left[\mathbf{w}\right]^2_n$. Thus, the same technique by fixing $\left[\mathbf{v}\right]^2_n + \left[\mathbf{w}\right]^2_n = P_n$ does not work in this case. Here, we present a slightly different approach 
to obtain simpler alternative expressions for $\text{SINR}^{\text{(2-branch)}}_{\text{B}}$ and $\text{SINR}^{\text{(2-branch)}}_{\text{E}}$. For this purpose, let $\widetilde{\mathbf{R}}_d(r)$, $\widetilde{\mathbf{R}}_z(r)$, $\widetilde{\pmb{\sigma}}_{\text{clip}, d}(r)$, $\widetilde{\pmb{\sigma}}_{\text{clip}, z}(r)$, $\widetilde{\sigma}^{2^{\text{(2-branch)}}}_{\text{B, norm}}(r)$, and  $\widetilde{\sigma}^{2^{\text{(2-branch)}}}_{\text{E, norm}}(r)$ be the values of $\mathbf{R}_d$, $\mathbf{R}_z$, $\pmb{\sigma}_{\text{clip}, d}$, $\pmb{\sigma}_{\text{clip}, z}$, $\sigma^{2^{\text{(2-branch)}}}_{\text{B, norm}}$, and  $\sigma^{2^{\text{(2-branch)}}}_{\text{E, norm}}$ at $\left|\left[\mathbf{v}\right]_n\right| = \sqrt{rP_n}$ and $\left|\left[\mathbf{w}\right]_n\right| = \sqrt{(1-r)P_n}$ ($\forall n = 1, 2, ..., N_T$). Here, $r \in (0, ~1)$ is a fixed parameter, which controls the power allocations for the information-bearing and AN signals. Then, denote $\widetilde{\text{SINR}}^{\text{(2-branch)}}_{\text{B}}(r)$ and $\widetilde{\text{SINR}}^{\text{(2-branch)}}_{\text{E}}(r)$ as the values of ${\text{SINR}}^{\text{(2-branch)}}_{\text{B}}$ and ${\text{SINR}}^{\text{(2-branch)}}_{\text{E}}$ at $\mathbf{R}_d = \widetilde{\mathbf{R}}_d(r)$, $\mathbf{R}_z = \widetilde{\mathbf{R}}_z(r)$,   ${\pmb{\sigma}}_{\text{clip}, d} = \widetilde{\pmb{\sigma}}_{\text{clip}, d}(r)$, ${\pmb{\sigma}}_{\text{clip}, z} = \widetilde{\pmb{\sigma}}_{\text{clip}, z}(r)$, ${\sigma}^{2^{\text{(2-branch)}}}_{\text{B, norm}} = \widetilde{\sigma}^{2^{\text{(2-branch)}}}_{\text{B, norm}}(r)$, and ${\sigma}^{2^{\text{(2-branch)}}}_{\text{E, norm}} = \widetilde{\sigma}^{2^{\text{(2-branch)}}}_{\text{E, norm}}(r)$. We then consider the following design problem
\begin{subequations}
\label{OptProb10}
    \begin{alignat}{2}
        &\underset{\mathbf{v}, \mathbf{w}}{\text{maximize}} & \hspace{2mm} & \widetilde{\text{SINR}}^{\text{(2-branch)}}_{\text{B}}(r) \label{obj10}\\
        &\text{subject to }  &  & \nonumber \\
        & & & \widetilde{\text{SINR}}^{\text{(2-branch)}}_{\text{E}}(r)  \leq \lambda, \label{constraint101}\\
        & & & \left[\mathbf{v}\right]_n^2 + \left[\mathbf{w}\right]_n^2 \leq P_n,~~ \forall n=1, 2, ..., N_T, \label{constraint102}
    \end{alignat}
\end{subequations}
which can be solved using the same approach presented for the one-branch AN transmission scheme. For a given $r$, the values of $\text{SINR}^{\text{(2-branch)}}_{\text{B}}$ and $\text{SINR}^{\text{(2-branch)}}_{\text{E}}$ corresponding to the solution $\mathbf{v}^*$ and $\mathbf{w}^*$ to \eqref{OptProb10} are denoted as $\text{SINR}^{*^\text{(2-branch)}}_{\text{B}}(r)$ and $\text{SINR}^{*^\text{(2-branch)}}_{\text{E}}(r)$, respectively. $\widetilde{\text{SINR}}^{*^{\text{(2-branch)}}}_{\text{B}}$ and $\widetilde{\text{SINR}}^{*^{\text{(2-branch)}}}_{\text{E}}$ are defined using the same manner. 

An issue of this design approach is the choice of $r$, which influences the values of $\widetilde{\mathbf{R}}_d(r)$, $\widetilde{\mathbf{R}}_z(r)$, $\widetilde{\pmb{\sigma}}_{\text{clip}, d}(r)$, $\widetilde{\pmb{\sigma}}_{\text{clip}, z}(r)$, $\widetilde{\sigma}^{2^{\text{(2-branch)}}}_{\text{B, norm}}(r)$, and  $\widetilde{\sigma}^{2^{\text{(2-branch)}}}_{\text{E, norm}}(r)$, hence the optimal values  $\widetilde{\text{SINR}}_{\text{B}}^{*^{\text{(2-branch)}}}(r)$, $\widetilde{\text{SINR}}_{\text{E}}^{*^{\text{(2-branch)}}}(r)$ to \eqref{OptProb10} and the resulting ${\text{SINR}}_{\text{B}}^{*^{\text{(2-branch)}}}(r)$, ${\text{SINR}}_{\text{E}}^{*^{\text{(2-branch)}}}(r)$. Interestingly, numerical results show that while $r$ has a considerable impact on $\widetilde{\text{SINR}}^{*^{\text{(2-branch)}}}_{\text{B}}(r)$ , it insignificantly affects  $\widetilde{\text{SINR}}^{*^{\text{(2-branch)}}}_{\text{E}}(r)$, $\text{SINR}_{\text{B}}^{*^{\text{(2-branch)}}}(r)$, and $\text{SINR}_{\text{E}}^{*^{\text{(2-branch)}}}(r)$ under the considered parameter settings. This is indeed a desirable feature of the proposed approach.  
\subsection{Unknown $\mathbf{h}_{\rm{E}}$}
The average of Eve's SINR when $\mathbf{h}_{\text{E}}$ is unknown to the transmitter is given as follows
\begin{align}
     \overline{\text{SINR}}^{(\text{2-branch})}_{\text{E}} = \mathbb{E}_{\mathbf{h}_{\text{E}}}\left[\frac{\left(n_{c,d}\left(\mathbf{h}_{\text{E}}\odot\mathbf{R}_{d}\right)^T{\mathbf{v}}\right)^2}{ \left(n_{c,d}\mathbf{h}^T_{\text{E}}\pmb{\sigma}_{\text{clip}, d}\right)^2 + n_{c,z}^2\left(\left(\left(\mathbf{h}_{\text{E}}\odot\mathbf{R}_{z}\right)^T{\mathbf{w}}\right)^2 + \left(\mathbf{h}^T_{\text{E}}\pmb{\sigma}_{\text{clip}, z}\right)^2\right) + {\sigma}^{2^{\text{(2-branch)}}}_{\text{E, norm}}}\right].
 \end{align}
Similar to the one-branch AN transmission scheme, we consider an alternative expression to $\overline{\text{SINR}}^{(\text{2-branch})}_{\text{E}}$, which is defined by 
\begin{align}
    \lambda^{\text{(2-branch)}}_{\text{E}} & =  \frac{\mathbb{E}_{\mathbf{h}_{\text{E}}}\left[{\left(n_{c,d}\left(\mathbf{h}_{\text{E}}\odot\mathbf{R}_d\right)^T{\mathbf{v}}\right)^2}\right]}{\mathbb{E}_{\mathbf{h}_{\text{E}}}\left[ \left(n_{c,d}\mathbf{h}^T_{\text{E}}\pmb{\sigma}_{\text{clip}, d}\right)^2 + n_{c,z}^2\left(\left(\left(\mathbf{h}_{\text{E}}\odot\mathbf{R}_z\right)^T{\mathbf{w}}\right)^2 + \left(\mathbf{h}^T_{\text{E}}\pmb{\sigma}_{\text{clip}, z}\right)^2\right) + {\sigma}^{2^{\text{(2-branch)}}}_{\text{E, norm}}\right]} \nonumber \\ 
    & = \frac{n_{c, d}^2\mathbf{v}^T\left(\left(\mathbf{R}_d\mathbf{R}^T_d\right)\odot\overline{\mathbf{H}}_{\text{E}}\right)\mathbf{v}}{n^2_{c,d}\pmb{\sigma}^T_{\text{clip}, d}\overline{\mathbf{H}}_{\text{E}}\pmb{\sigma}_{\text{clip}, d} + n^2_{c,z}\pmb{\sigma}^T_{\text{clip}, z}\overline{\mathbf{H}}_{\text{E}}\pmb{\sigma}_{\text{clip}, z} + n^2_{c, z}\mathbf{w}^T\left(\left(\mathbf{R}z\mathbf{R}^T_z\right)\odot\overline{\mathbf{H}}_{\text{E}}\right)\mathbf{w} + {\overline{\sigma}}^{2^{\text{(2-branch)}}}_{\text{E, norm}}}.
        \label{EveSINR4}
\end{align}
Similar to the case of known $\mathbf{h}_{\text{E}}$, denote $\widetilde{\lambda}^{\text{(2-branch)}}_{\text{E}}(r)$ as the value of $\lambda^{\text{(2-branch)}}_{\text{E}}$ at $\mathbf{R}_d = \widetilde{\mathbf{R}}_d(r)$, $\mathbf{R}_z = \widetilde{\mathbf{R}}_z(r)$,   ${\pmb{\sigma}}_{\text{clip}, d} = \widetilde{\pmb{\sigma}}_{\text{clip}, d}(r)$, ${\pmb{\sigma}}_{\text{clip}, z} = \widetilde{\pmb{\sigma}}_{\text{clip}, z}(r)$, ${\sigma}^{2^{\text{(2-branch)}}}_{\text{B, norm}} = \widetilde{\sigma}^{2^{\text{(2-branch)}}}_{\text{B, norm}}(r)$, and ${\sigma}^{2^{\text{(2-branch)}}}_{\text{E, norm}} = \widetilde{\sigma}^{2^{\text{(2-branch)}}}_{\text{E, norm}}(r)$. This gives rise to the following problem
\begin{subequations}
\label{OptProb11}
    \begin{alignat}{2}
        &\underset{\mathbf{v}, \mathbf{w}}{\text{maximize}} & \hspace{2mm} & \widetilde{\text{SINR}}^{\text{(2-branch)}}_{\text{B}}(r) \label{obj11}\\
        &\text{subject to }  &  & \nonumber \\
        & & & \widetilde{\lambda}^{\text{(2-branch)}}_{\text{E}}(r)  \leq \lambda, \label{constraint111}\\
        & & & \left[\mathbf{v}\right]_n^2 + \left[\mathbf{w}\right]_n^2 \leq P_n,~~ \forall n=1, 2, ..., N_T, \label{constraint112}
    \end{alignat}
\end{subequations}
which, again, can be solved using the above-mentioned procedure. Then, for a given $r$, let $\widetilde{\text{SINR}}^{*^{\text{(2-branch)}}}_{\text{B}}(r)$, $\widetilde{\lambda}^{*^{\text{(2-branch)}}}_{\text{E}}(r)$,  
$\text{SINR}^{*^{(\text{2-branch})}}_{\text{B}}(r)$,  $\lambda^{*^\text{(2-branch)}}_{\text{E}}(r)$, $\overline{\text{SINR}}^{*^{\text{(2-branch)}}}_{\text{E}}(r)$ are the values of  
$\widetilde{\text{SINR}}^{{\text{(2-branch)}}}_{\text{B}}(r)$, $\widetilde{\lambda}^{{\text{(2-branch)}}}_{\text{E}}(r)$,  
$\text{SINR}^{{(\text{2-branch})}}_{\text{B}}(r)$,  $\lambda^{\text{(2-branch)}}_{\text{E}}(r)$, $\overline{\text{SINR}}^{{\text{(2-branch)}}}_{\text{E}}(r)$ at the obtained optimal solution, respectively. 
\begin{table}[H]
\caption{System Parameters} 
\centering 
\begin{tabular}{l l l} 
\midrule
\multicolumn{2}{c}{\bf{Room and LED configurations}} \\
\midrule 
Room dimension, $d_L \times d_W \times d_H$ &  5 m$\times$ 5 m$\times$ 3 m  \\
 \midrule
 Height of the receiver plan, $d_R$ & 0.5 m \\
 \midrule
 LED luminary positions & Luminary 1 : $(-\sqrt{2}$, $-\sqrt{2}, 3)$   ~ ~&Luminary 2 : $(\sqrt{2}$, $-\sqrt{2}, 3)$\\ & Luminary 3 : $(\sqrt{2}$, $\sqrt{2}, 3)$ ~& Luminary 4 : $-\sqrt{2}$, $\sqrt{2}, 3)$ \\
 \midrule     
 LED bandwidth, $B_{\text{mod}}$ & 20 MHz \\
 \midrule
 Number of chips in each luminary, $n_c$ & 24 \\
 \midrule     
 LED beam angle, $\phi$ & $120^\circ$  \\
 \midrule 
 LED conversion factor, $\eta$ & 0.44 W/A  \\
 \midrule
 $[I_{\text{min}}~~I_{\text{max}}]$ & $[0~\text{A}~~1~\text{A}]$ \\
 \midrule     
 \multicolumn{2}{c}{\bf{Receiver photodetectors}} \\
 \midrule     
 Active area, $A_r$ & 1 $\text{cm}^2$ \\ 
 \midrule     
 Responsivity, $\gamma$ & 0.54 A/W\\ 
 \midrule     
 Field of view (FoV), $\Psi$ & $60^\circ$\\ 
 \midrule     
 Optical filter gain, $T_s(\psi)$ & 1\\ 
 \midrule     
 Refractive index of the concentrator, $\kappa$ & 1.5\\ 
 \midrule 
 \multicolumn{2}{c}{\bf{Other parameters}} \\
 \midrule 
 Ambient light photocurrent, $\chi_{\text{amp}}$ & 10.93 $\text{A}/(\text{m}^2 \cdot \text{Sr}$)\\
\midrule 
Preamplifier noise current density, $i_{\text{amp}}$ & 5 $\text{pA}/\text{Hz}^{-1/2}$ \\
\midrule 
\end{tabular}
\label{table1}
\end{table} 
\section{Numerical Results and Discussions}
\label{sec:numerical-discussions}
In this section, simulation results are provided to illustrate the secrecy performances of the one-branch and two-branch AN designs presented in the previous section. Without otherwise noted, system parameters are given in Table I, and the geometrical configuration is illustrated in Fig.~\ref{Fig0}. Moreover, simulation results are obtained by averaging over 5000 randomly located positions of Bob and Eve. For the proposed AN transmission scheme, assume that each LED luminary is composed of 24 chips which are divided evenly into two groups (i.e., $n_{c, d} = n_{c, z} = 12$). 
\begin{figure}[ht]
    \centering
    \begin{subfigure}[b]{.6\textwidth}
        \includegraphics[width = .95\textwidth, height = 5.2cm]{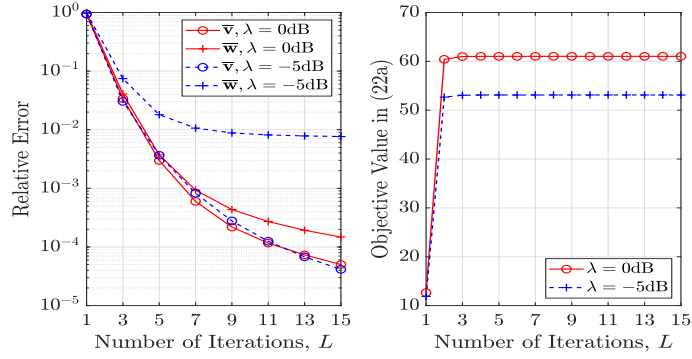}
        \caption{Known $\mathbf{h}_{\text{E}}$ - Problem \eqref{OptProb6}.}
        \label{convergenceKnownHe1}
    \end{subfigure}
    \begin{subfigure}[b]{.6\textwidth}
        \includegraphics[width = .95\textwidth, height = 5.2cm]{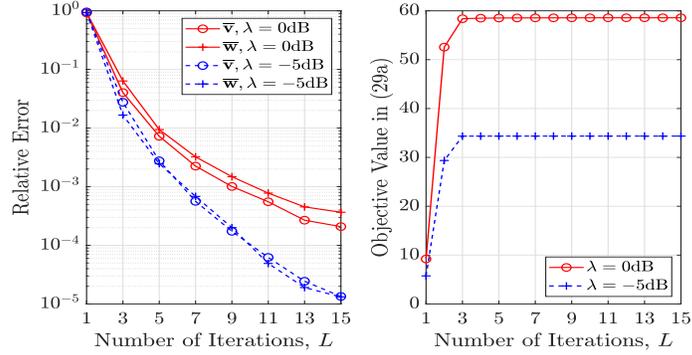}
        \caption{Unknown $\mathbf{h}_{\text{E}}$ - Problem \eqref{OptProb8-1}.}
        \label{convergenceUnknownHe1}
    \end{subfigure}
    \caption{Convergence behaviors of \textbf{Algorithm 1}.}
    \label{convergenceAlgo1}
\end{figure}

Firstly, Figs.~\ref{convergenceKnownHe1} and \ref{convergenceUnknownHe1} depict the convergence behaviors of  $\textbf{Algorithm 1}$ in solving \eqref{OptProb6} and \eqref{OptProb8-1} with respect to the number of iterations $L$ considering that $\lambda = 0$ and $-5$ dB. It is shown in both cases that, on average, the objective functions in \eqref{obj6} and \eqref{obj8-1} approach their optimal values when $L\approx 3$, which corresponds to relative errors of $\overline{\mathbf{v}}$ and $\overline{\mathbf{w}}$ being greater than $10^{-2}$, respectively. Accordingly, the error tolerance $\epsilon = 10^{-2}$ and the maximum number of iteration $L_{\text{max}} = 10$ can be chosen to guarantee a satisfactory convergence of \eqref{obj6} and \eqref{obj8-1}.

\begin{figure*}[ht]
    \centering
    \begin{subfigure}[b]{\textwidth}
        \centering
        \includegraphics[width = .85\textwidth, height = 4.8cm]{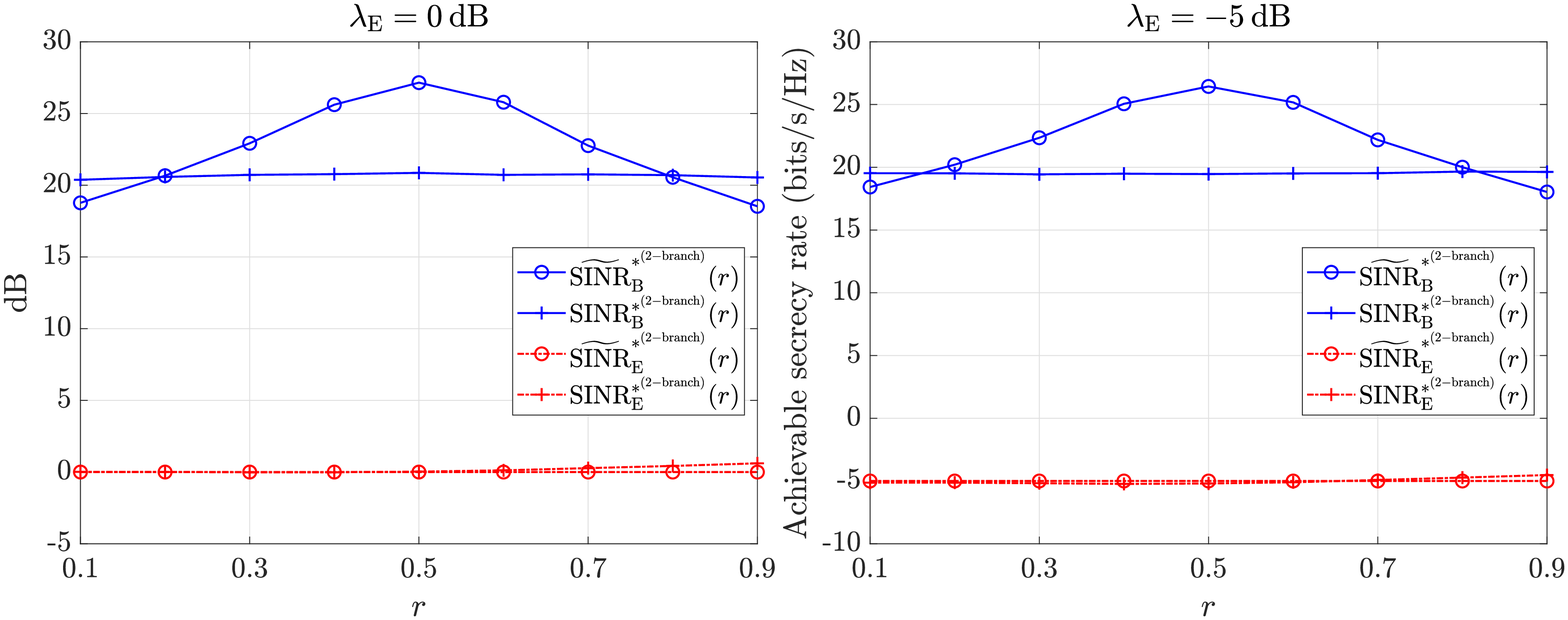}
        \caption{Known $\mathbf{h}_{\text{E}}$.}
        \label{KnownHe:Secrecy_vs_R}
    \end{subfigure}
    \begin{subfigure}[b]{\textwidth}
        \centering
        \includegraphics[width = .85\textwidth, height = 4.8cm]{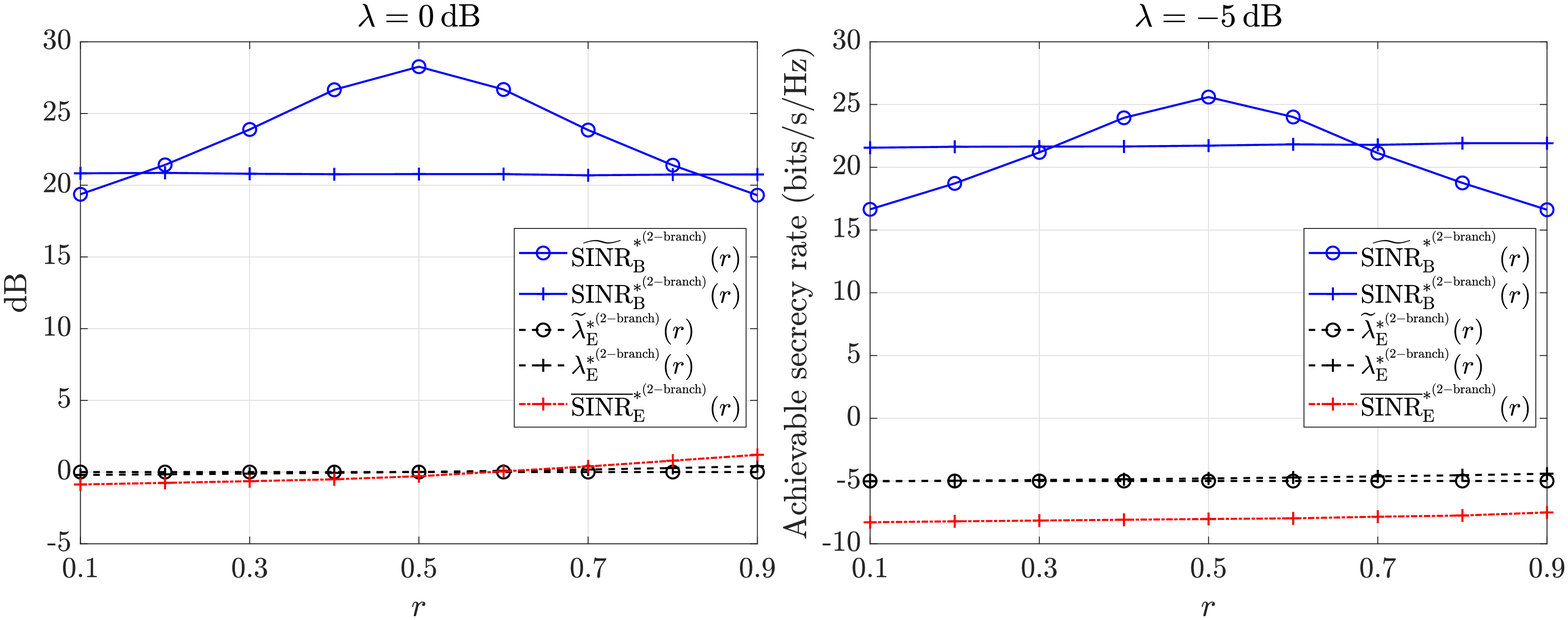}
        \caption{Unknown $\mathbf{h}_{\text{E}}$.}
        \label{UnknownHe:Secrecy_vs_R}
    \end{subfigure}
    \caption{Secrecy performance versus $r$ for different thresholds on Eve's SINR, $\sigma_P = 0.25$A.}
    \label{Secrecy_vs_R}
\end{figure*}

Next, the impact of selecting $r$ on the secrecy performance of the proposed AN transmission scheme when $\sigma_P = 0.25$ A (i.e, $I_{\text{DC}} = 0.5$ A) is shown in  Figs. \ref{KnownHe:Secrecy_vs_R}  and \ref{UnknownHe:Secrecy_vs_R} for the case of known and unknown $\mathbf{h}_{\text{E}}$, respectively. Regarding the solution to \eqref{OptProb10} in the case of known $\mathbf{h}_{\text{E}}$, 
it is noticed that while $\widetilde{\text{SINR}}^{*^{\text{(2-branch)}}}_{\text{B}}(r)$ changes drastically with $r$ and peaks at $r = 0.5$, $\widetilde{\text{SINR}}^{*^{\text{(2-branch)}}}_{\text{E}}(r)$ stays unchanged at the predefined threshold $\lambda$. An intuitive explanation for this can be given as follows.  As long as the constraint in \eqref{constraint102} is satisfied, in order for $\widetilde{\text{SINR}}^{\text{(2-branch)}}_{\text{B}}(r)$ to be  maximized, the degrees of freedom in choosing $\mathbf{v}$ and $\mathbf{w}$ should be as high as possible, which implies that   $\widetilde{\text{SINR}}^{{\text{(2-branch)}}}_{\text{E}}(r)$ tend to archive its maximum allowable value of $\lambda$ at the optimal solution. 
The same behavior is also observed in the case of unknown $\mathbf{h}_{\text{E}}$ as seen in Fig. \ref{UnknownHe:Secrecy_vs_R}. Interestingly, though the selection of $r$ significantly influences $\widetilde{\text{SINR}}^{*^{\text{(2-branch)}}}_{\text{B}}$, it has a very little impact on $\text{SINR}^{*^{\text{(2-branch)}}}_{\text{B}}$, $\text{SINR}^{*^{\text{(2-branch)}}}_{\text{E}}$(in the case of unknown $\mathbf{h}_{\text{E}}$), $\lambda^{*^{\text{(2-branch)}}}_{\text{E}}$, and $\overline{\text{SINR}}^{*^{\text{(2-branch)}}}_{\text{E}}$(in the case of known $\mathbf{h}_{\text{E}}$). This is indeed desirable as the secrecy performance of the proposed method can be evaluated independently of $r$. Therefore, in the subsequent simulations, $r = 0.5$ is chosen and the index $r$ is omitted for the sake of conciseness.         
\begin{figure*}[ht]
    \centering
    \begin{subfigure}[b]{\textwidth}
        \centering
        \includegraphics[width = .85\textwidth, height = 4.8cm]{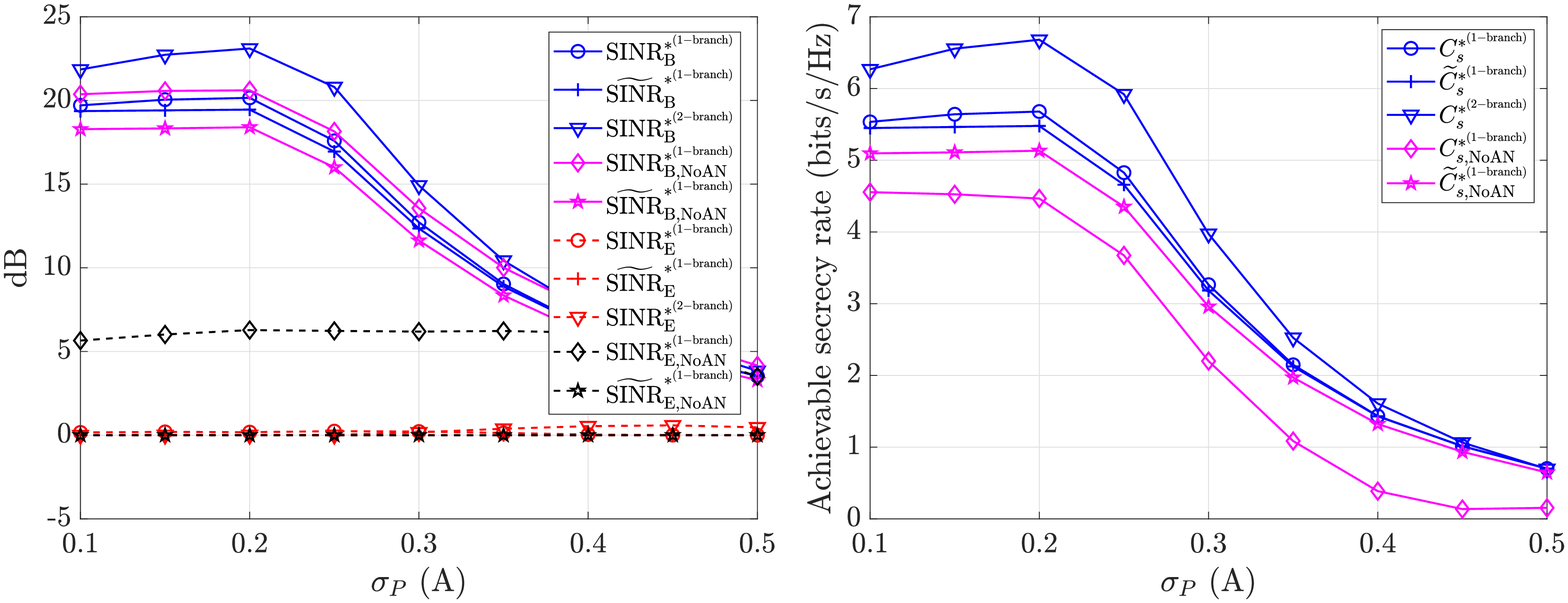}
        \caption{$\lambda = 0$ dB.}
        \label{KnownHe:SecrecyPower1}
    \end{subfigure}
    \begin{subfigure}[b]{\textwidth}
        \centering
        \includegraphics[width = .85\textwidth, height = 4.8cm]{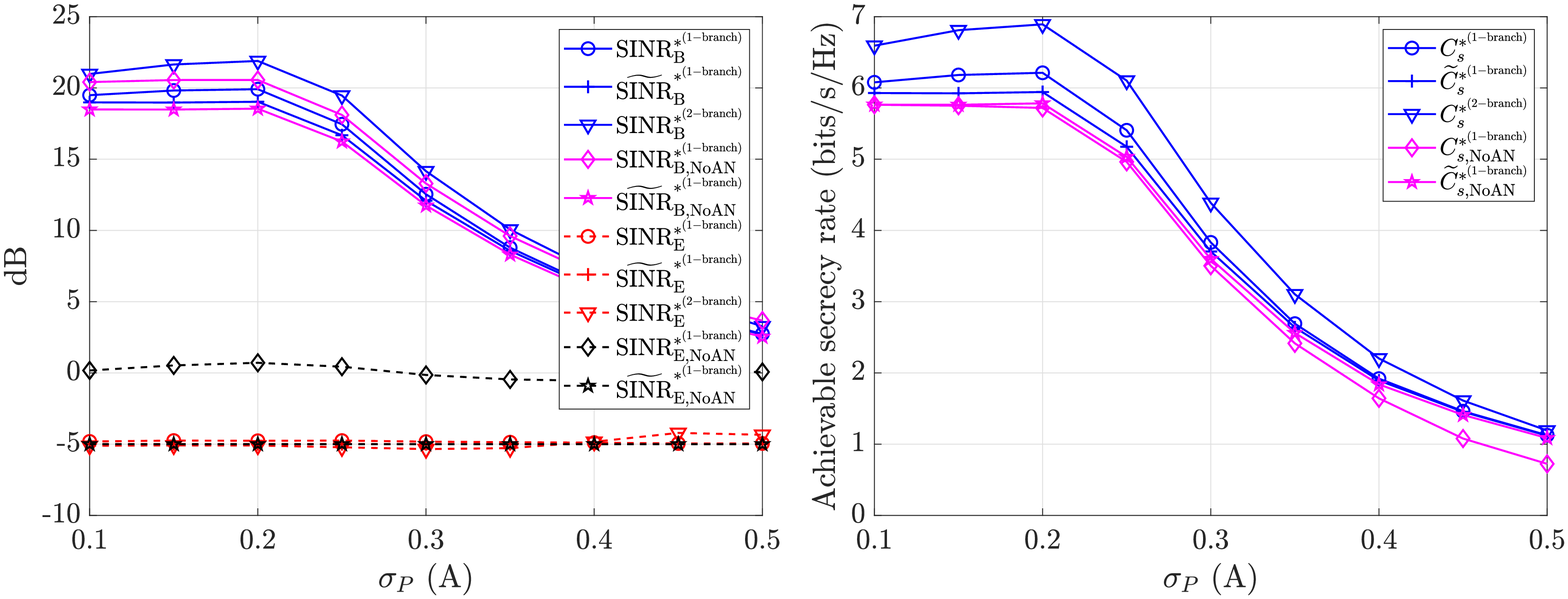}
        \caption{$\lambda = -5$ dB.}
        \label{KnownHe:SecrecyPower2}
    \end{subfigure}
    \caption{Secrecy performance versus $\sigma_{P}$ for different thresholds on Eve's SINR: Known $\mathbf{h}_{\text{E}}$.}
    \label{KnownHe:Secrecy_vs_Power}
\end{figure*}

With the above setting and in the case of known $\mathbf{h}_{\text{E}}$, the secrecy performances of the one-branch and two-branch AN transmission schemes are displayed as a function of $\sigma_P$ in Figs. \ref{KnownHe:SecrecyPower1} and \ref{KnownHe:SecrecyPower2} for $\lambda = 0$ dB and $-5$ dB, respectively. Recall that $\sigma_P$ denotes the maximum allowable standard deviation of the combined AN and the information-bearing signal. To highlight the benefit of using AN, secrecy performances in the case of no-AN (i.e., $\mathbf{w} = \mathbf{0}$) represented by $\text{SINR}^{*^{\text{(1-branch)}}}_{\text{B, NoAN}}$, $\widetilde{\text{SINR}}^{*^{\text{(1-branch)}}}_{\text{B, NoAN}}$, $\text{SINR}^{*^{\text{(1-branch)}}}_{\text{E, NoAN}}$, $\widetilde{\text{SINR}}^{*^{\text{(1-branch)}}}_{\text{E, NoAN}}$, $C^{*^{\text{(1-branch)}}}_{s, \text{NoAN}}$, $\widetilde{C}^{*^{\text{(1-branch)}}}_{s, \text{NoAN}}$ are also shown for comparison. Note that $C^{*^{\text{(1-branch)}}}_{s, \text{NoAN}}$ and $\widetilde{C}^{*^{\text{(1-branch)}}}_{s, \text{NoAN}}$ are defined in the same manner as $C^{*^{\text{(1-branch)}}}_s$ and $\widetilde{C}^{*^{\text{(1-branch)}}}_s$. Firstly, the left sub-figures compare the AN and no-AN approaches in terms of Bob's and Eve's SINR performances. In the AN scheme,  notice that the differences between $\text{SINR}^{*^{\text{(1-branch)}}}_{\text{B}}$ and $\widetilde{\text{SINR}}^{*^{\text{(1-branch)}}}_{\text{B}}$ are small, especially at high values of $I_{\text{DC}}$. More importantly, the differences between $\text{SINR}^{*^{\text{(1-branch)}}}_{\text{E}}$ and $\widetilde{\text{SINR}}^{*^{\text{(1-branch)}}}_{\text{E}}$ are seen to be negligible. These two observations confirm the approximation of the proposed sub-optimal approach presented in Sec.~\ref{KnownHe_AN}. In the case of no-AN scheme, the gap between $\text{SINR}^{*^{\text{(1-branch)}}}_{\text{B, NoAN}}$ and $\widetilde{\text{SINR}}^{*^{\text{(1-branch)}}}_{\text{B, NoAN}}$ are more clear but still near 3~dB. Nonetheless, there are considerable gaps between  $\text{SINR}^{*^{\text{(1-branch)}}}_{\text{E, NoAN}}$ and $\widetilde{\text{SINR}}^{*^{\text{(1-branch)}}}_{\text{E, NoAN}}$, where  Eve's SINRs are about 5 dB higher than the results obtained from using the approximate expression. The superiority of the proposed two-branch AN transmission scheme is also clearly shown as it is observed that $\text{SINR}^{*^{\text{(2-branch)}}}_{\text{E}}$ is negligibly higher than $\text{SINR}^{*^{\text{(1-branch)}}}_{\text{E}}$ while $\text{SINR}^{*^{\text{(2-branch)}}}_{\text{B}}$ is about 3 dB and 2.5 dB better than $\text{SINR}^{*^{\text{(1-branch)}}}_{\text{B}}$ at $\lambda = 0$ dB and $-5$ dB, respectively. For a better illustration of the secrecy performance, the right sub-figures compare the three AN transmission schemes in terms of the achievable secrecy rate, where $C^{*^{\text{(2-branch)}}}_s = \log_2\left(1 + \text{SINR}^{*^{\text{(2-branch)}}}_{\text{B}
}\right) - \log_2\left(1 + \text{SINR}^{*^{\text{(2-branch)}}}_{\text{E}
}\right)$. Then, improvements in the achievable secrecy rate are clearly visualized. For example, at $\sigma_{P} = 0.25$A, compared with the no-AN scheme, the one-branch AN approach improves the achievable secrecy rate by about 1.2 and 0.4 bits/s/Hz when $\lambda = 0$ and $-5$dB, respectively. On the other hand, the two-branch AN transmission scheme outperforms the one-branch one by 1 and 0.7 bits/s/Hz. Moreover,  observe that at low $\sigma_{P}$ values (i.e., below 0.2 A), the achievable secrecy rate stays nearly unchanged. This arises because, for low $\sigma_{P}$, the impact of increasing $\sigma_{P}$ is balanced by the clipping distortion. However, when $\sigma_{P}> 0.2$ A, the impact of the clipping distortion becomes dominant, resulting in a dramatic degradation of the achievable secrecy rate. 
\begin{figure*}[ht]
    \centering
    \begin{subfigure}[b]{\textwidth}
        \centering
        \includegraphics[width = .85\textwidth, height = 4.8cm]{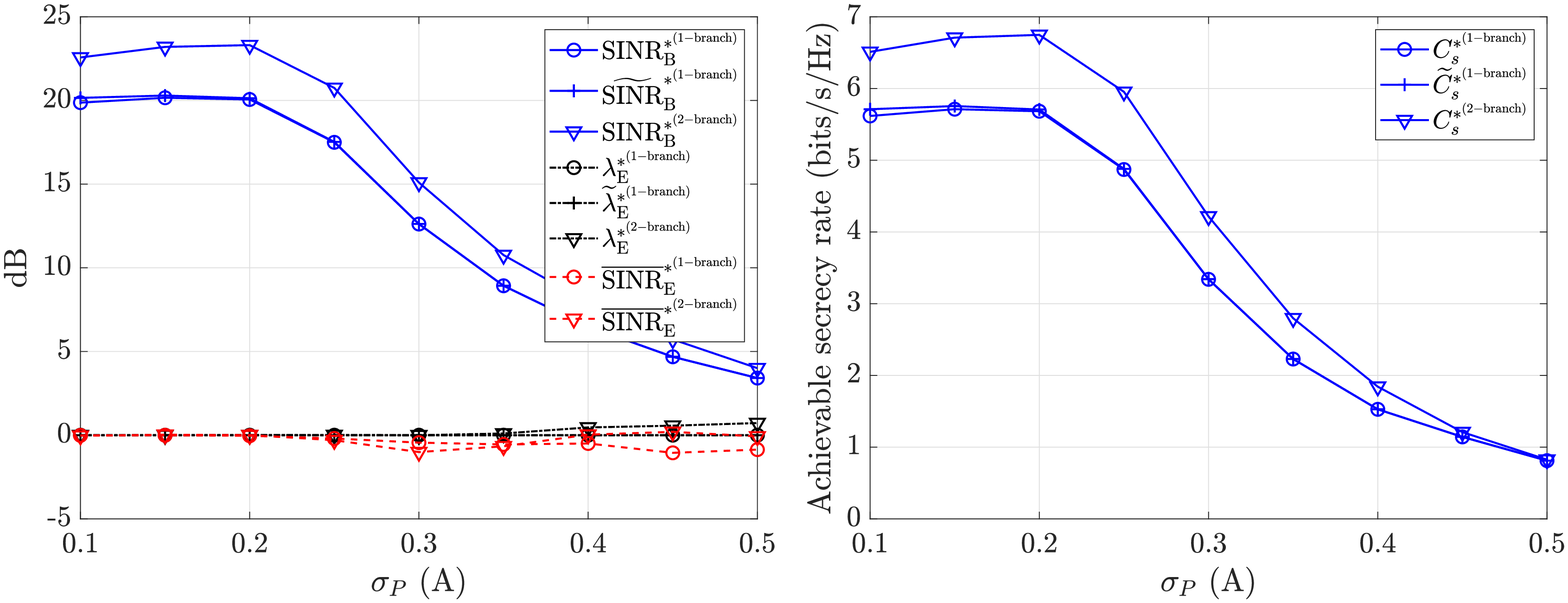}
        \caption{$\lambda = 0$ dB.}
        \label{UnknownHe:SecrecyPower1}
    \end{subfigure}
    \begin{subfigure}[b]{\textwidth}
        \centering
        \includegraphics[width = .85\textwidth, height = 4.8cm]{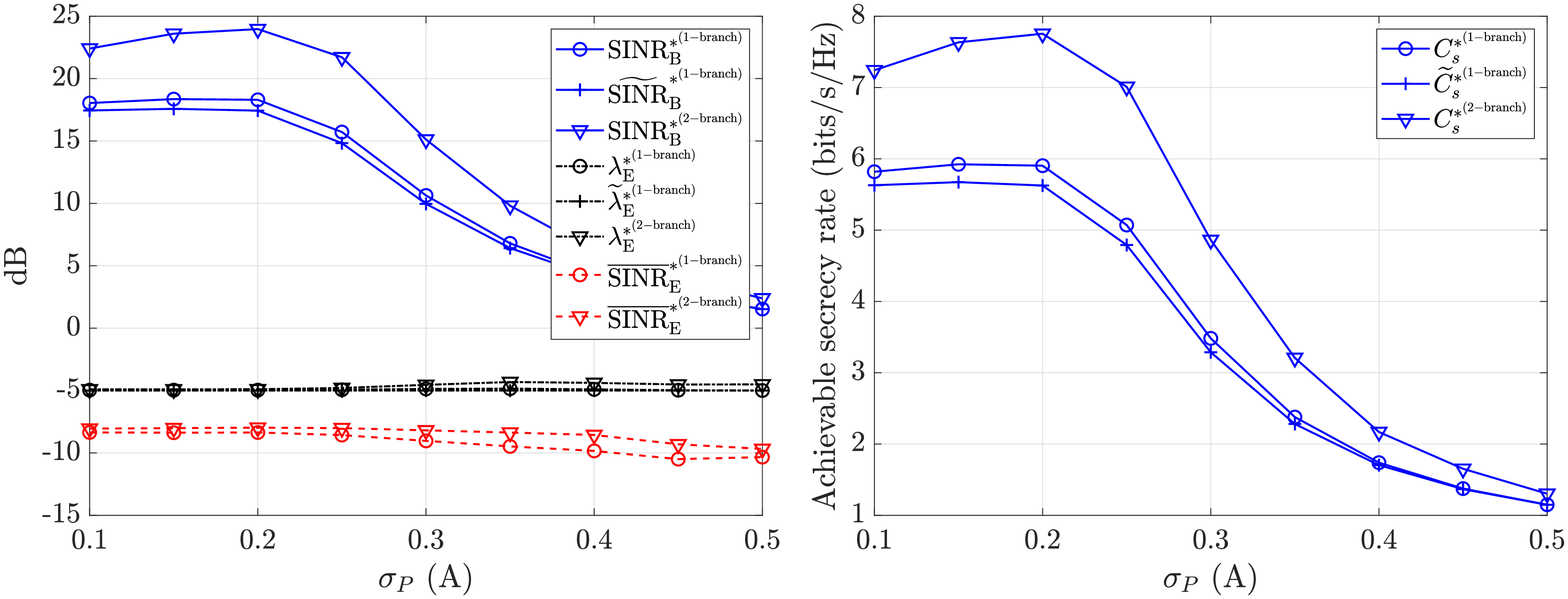}
        \caption{$\lambda = -5$ dB.}
        \label{UnknownHe:SecrecyPower2}
    \end{subfigure}
    \caption{Secrecy performance versus $\sigma_{P}$ for different thresholds on Eve's SINR: Unknown $\mathbf{h}_{\text{E}}$.}
    \label{UnknownHe:Secrecy_vs_Power}
\end{figure*}

For the case of unknown $\mathbf{h}_{\text{E}}$, the secrecy performances  with respect to $\sigma_P$ are illustrated in Figs. \ref{UnknownHe:SecrecyPower1} and \ref{UnknownHe:SecrecyPower2}. Similar to the case of known $\mathbf{h}_{\text{E}}$, the differences between $\text{SINR}^{*^{\text{(1-branch)}}}_{\text{B}}$ and $\widetilde{\text{SINR}}^{*^{\text{(1-branch)}}}_{\text{B}}$ as well as between $\lambda^{*^{\text{(1-branch)}}}_{\text{E}}$ and $\widetilde{\lambda}^{*^{\text{(1-branch)}}}_{\text{E}}$ are negligible, which, again, validates the usefulness of the proposed sub-optimal design approach. More importantly, the resulting actual average Eve's SINR (i.e., $\overline{\text{SINR}}^{*^{\text{(1-branch)}}}_{\text{E}}$) is seen to be lower than the alternative term $\lambda^{*^{\text{(1-branch)}}}_{\text{E}}$ in both cases of $\lambda$. While it is difficult to give an analytical comparison between  $\overline{\text{SINR}}^{*^{\text{(1-branch)}}}_{\text{E}}$ and $\lambda^{*^{\text{(1-branch)}}}_{\text{E}}$, our simulation results under various values of the threshold $\lambda$ show that $\overline{\text{SINR}}^{*^{\text{(1-branch)}}}_{\text{E}}$ is smaller than $\lambda^{*^{\text{(1-branch)}}}_{\text{E}}$ when $\lambda$ is set to be sufficiently small, which is in fact preferable in practical systems as to degrade the eavesdropped signal quality. In this scenario, the achievable secrecy rates are given by $C^{*^{\text{(1-branch)}}}_s = \log_2\left(1 + \text{SINR}^{*^{\text{(1-branch)}}}_{\text{B}}\right) - \log_2\left(1 + \overline{\text{SINR}}^{*^{\text{(1-branch)}}}_{\text{E}}\right)$, $\widetilde{C}^{*^{\text{(1-branch)}}}_s = \log_2\left(1 + \widetilde{\text{SINR}}^{*^{\text{(1-branch)}}}_{\text{B}}\right) - \log_2\left(1 + \overline{\text{SINR}}^{*^{\text{(1-branch)}}}_{\text{E}}\right)$, and $C^{*^{\text{(2-branch)}}}_s = \log_2\left(1 + \text{SINR}^{*^{\text{(2-branch)}}}_{\text{B}}\right) - \log_2\left(1+\overline{\text{SINR}}^{*^{\text{(2-branch)}}}_{\text{E}}\right)$. Here, the superiority of the two-branch AN transmission scheme over the one-branch one is also demonstrated. For instance, at $\sigma_P = 0.25$ A, improvements of about 1 and 2 bits/s/Hz are attained when $\lambda = 0$ dB and $-5$ dB, respectively. Also, the impact of clipping distortion on the secrecy performances as $\sigma_P$ increases is qualitatively the same as in the case of known $\mathbf{h}_{\text{E}}$. It is worth mentioning that the achievable secrecy rate of the two-branch transmission scheme in the case of unknown $\mathbf{h}_{\text{E}}$ is almost the same as that in the case of known $\mathbf{h}_{\text{E}}$ when $\lambda = 0$ dB and even better when $\lambda = -5$ dB. This may seem counterintuitive but can be seen as a consequence of the proposed solving approach (i.e., optimization over $\widetilde{\text{SINR}}^{\text{(2-branch)}}_{\text{B}}(r)$ and $\widetilde{\text{SINR}}^{\text{(2-branch)}}_{\text{E}}(r)$ instead of ${\text{SINR}}^{\text{(2-branch)}}_{\text{B}}$ and ${\text{SINR}}^{\text{(2-branch)}}_{\text{E}}$).        
\begin{figure*}[ht]
    \centering
    \begin{subfigure}[b]{\textwidth}
    \centering
    \includegraphics[width = 0.85\textwidth, height = 4.8cm]{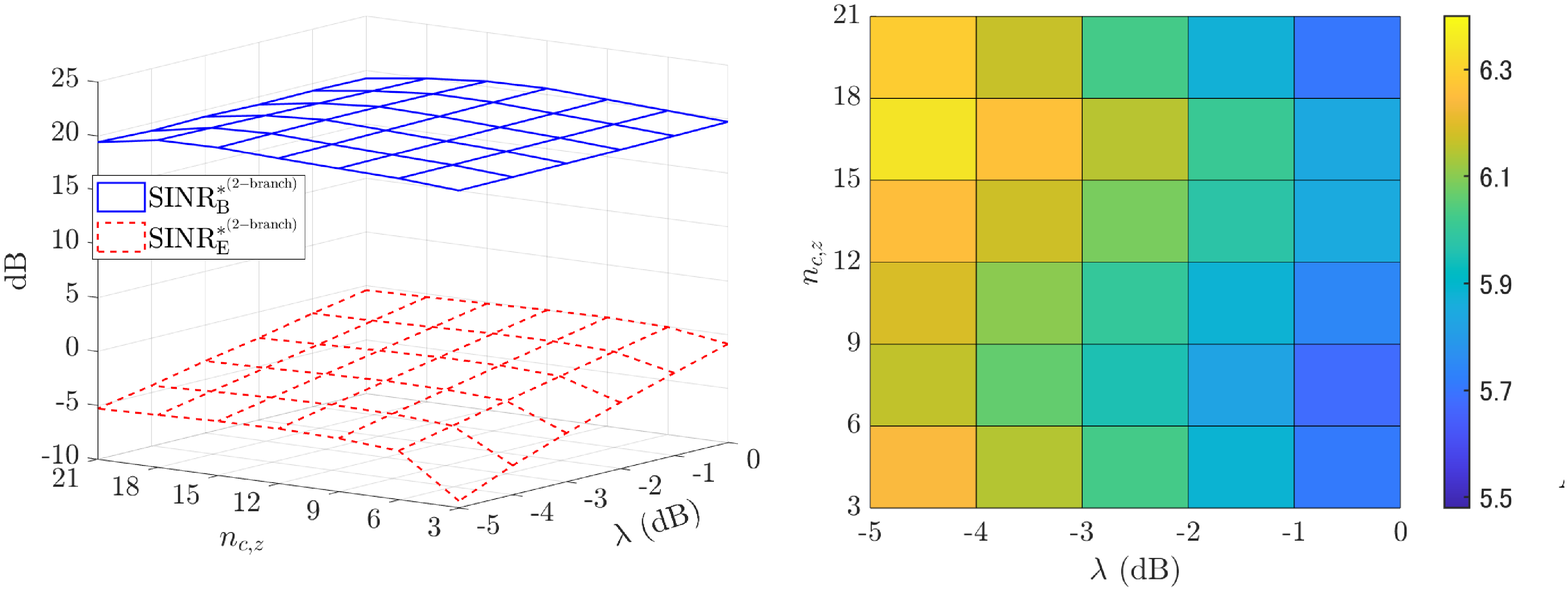}
    \caption{Known $\mathbf{h}_{\text{E}}$.}
    \label{SecrecyChipRatio_and_SINR1}
    \end{subfigure}
    \begin{subfigure}[b]{\textwidth}
    \centering
    \includegraphics[width = 0.85\textwidth, height = 4.8cm]{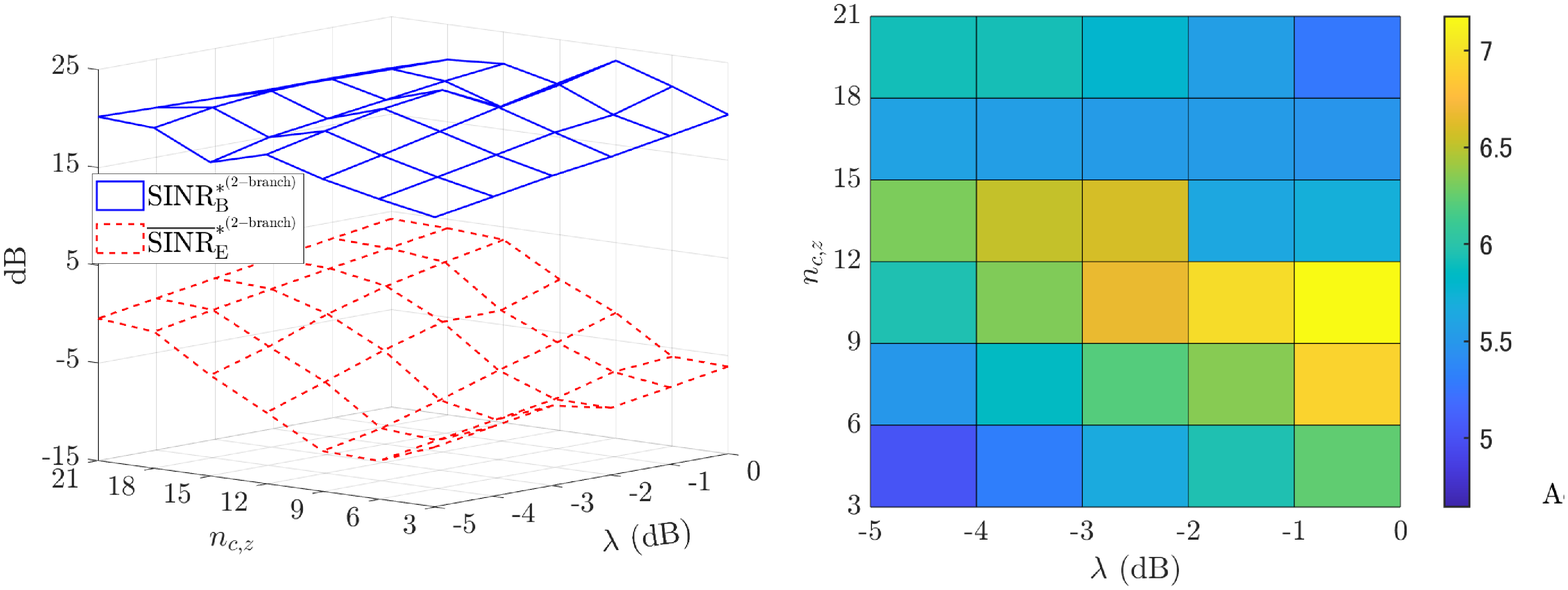}
    \caption{Unknown $\mathbf{h}_{\text{E}}$.}
    \label{SecrecyChipRatio_and_SINR2}
    \end{subfigure}
    \caption{Secrecy performance versus $\lambda$ and $n_{c, z}$ with $I_{\text{DC}} = 0.5$ A.}
\end{figure*}

A feature of the proposed two-branch AN transmission scheme is that it can flexibly adjust the number of chips in each group. Implementation of such an adjusting scheme in practical systems might add some complexity to the circuit design. Nonetheless, as system settings (e.g., the threshold $\lambda$) can be varied according to the required secrecy level, configuring the number of chips in each group can be beneficial to further improve the secrecy performances. For this purpose, we show in Figs. \ref{SecrecyChipRatio_and_SINR1} and \ref{SecrecyChipRatio_and_SINR2} the secrecy performances in accordance with the number of chips in the AN group $n_{c, z}$ and the threshold $\lambda$. Bob's position is fixed at $[-0.85,~-0.25,~0.5]$ and in the case of known $\mathbf{h}_{\text{E}}$, Eve's position is fixed at $[2.25,~1.85,~0.5]$. It is observed in both figures that   $\text{SINR}^{*^{\text{(2-branch)}}}_{\text{E}}$ and $\overline{\text{SINR}}^{*^{\text{(2-branch)}}}_{\text{E}}$ tend to increase as $n_{c, z}$ increases. This might seem counter-intuitive, yet can be explained by the fact that the majority of the power is allocated to the information-bearing signal (to ensure high Bob's SINRs). Thus, the impact of clipping distortion due to  the information-bearing signal is dominant. By increasing $n_{c, z}$ (i.e., reducing $n_{c, d}$), the distortion severity can be mitigated. However, when $n_{c, z}$ exceeds a certain value, the impact of AN and clipping distortion caused by the AN signal becomes considerable, which leads to a drop in   $\text{SINR}^{*^{\text{(2-branch)}}}_{\text{B}}$ and hence the achievable secrecy rate. As a result, one can choose an optimal value of $n_{c, z}$ to maximize the secrecy performance. For example, at $\lambda = 0$ dB, the optimal $n_{c, z}$ is around 15 in the case of known $\mathbf{h}_{\text{E}}$ while it is 10 in the case of unknown $\mathbf{h}_{\text{E}}$.              
\section{Conclusion}
In this paper, we have studied AN designs for PLS in VLC systems taking into account the impact of clipping distortion resulting from non-amplitude-constrained signals. Simulation results have shown  significant degradation in the secrecy performances due to the effect of clipping distortion and the benefit of using AN in improving the achievable secrecy rate compared with the no-AN scheme. Furthermore, by transmitting the information-bearing and AN signal through different groups of LED chips in each luminary, a novel AN transmission scheme has been proposed and shown to achieve considerable secrecy improvements.  Note that, to facilitate the analysis, the AN was assumed to be Gaussian, which, however, can incur severe clipping distortion. For future research directions, one may thus consider optimizing the AN's input distribution for further improvements of the secrecy performance. Another method for reducing the impact of clipping distortion is to employ more luminaries, which enables lower DC-bias to achieve the required illumination. It is then crucial to examine the effect of the number of luminaires on the secrecy performance. 
\label{sec:conclusion}
\bibliographystyle{IEEEtran}
\bibliography{references}
\end{document}